\documentclass[12pt]{article}
\usepackage{graphics}
\usepackage{amssymb}

\newcommand \be {\begin{equation}}
\newcommand \ee {\end{equation}}
\newcommand \ba {\begin{array}}
\newcommand \ea {\end{array}}
\newcommand \bea{\begin{eqnarray}}
\newcommand \eea{\end{eqnarray}}

\begin{document}

\begin{flushright}
YITP-SB-03-34 \\
\today
\end{flushright}

\vspace*{30mm}

\begin{center}
{\LARGE \bf Scattering Amplitudes in High Energy QCD}

\par\vspace*{20mm}\par

{\large \bf Tibor K\'ucs}

\bigskip

{\em C.N.\ Yang Institute for Theoretical Physics,
SUNY Stony Brook \\
Stony Brook, New York 11794 -- 3840, U.S.A.}

\end{center}
\vspace*{15mm}

\begin{abstract}

We develop a new systematic procedure for the
Regge limit in perturbative QCD to arbitrary logarithmic order.
The formalism relies on the IR structure and the gauge symmetry of the theory.
We identify leading regions in loop momentum space responsible for the
singular structure of the
amplitudes and perform power counting to determine the strength of these
divergences.
Using a factorization procedure introduced by Sen,
we derive a sum of convolutions in transverse momentum
space over soft and jet functions,
which approximate the amplitude up to power-suppressed corrections.
A set of evolution equations generalizing the BFKL equation and
controlling the high energy behavior of the amplitudes to arbitrary
logarithmic accuracy is derived. The general method is illustrated in the
case of leading logarithmic gluon reggeization
and BFKL equation.
\end{abstract}

\newpage

\section{Introduction}

The study of semihard processes within the framework of gauge quantum field
theories has a long history. For reviews see Refs.
\cite{lipatov}-\cite{ross}.
The defining feature of such processes is that they involve two or more hard
scales, compared to $\Lambda_{\rm QCD}$, which are strongly ordered relative
to each other.
The perturbative expansions of scattering amplitudes for these processes must
be resummed since they contain logarithmic enhancements
due to large ratios of the scales involved. One of the most important
examples is elastic $2 \rightarrow 2$
particle scattering in the Regge limit, $s \gg |t|$ (with $s$ and $t$ the
usual Mandelstam variables).
It is this process that we investigate in this paper.
We extend the techniques developed in Refs. \cite{sen83} and
\cite{jaroszewicz} and devise a new systematic method for evaluation of
QCD scattering amplitudes in the Regge limit to arbitrary logarithmic
accuracy.

The problem of the Regge limit in quantum field theory
was first tackled in the case of fermion exchange amplitude
within QED in Ref. \cite{gellMann}.
Here it was found that the positive signature amplitude takes a reggeized
form up to the two loop level in Leading Logarithmic
(LL) approximation. In Ref. \cite{wuMcCoy} the calculations were extended to
higher loops, and the imaginary
part of the Next-to-Leading Logarithms (NLL) was also obtained.
The analysis in Refs. \cite{gellMann} and \cite{wuMcCoy} was
performed in Feynman gauge.
It was realized in Ref. \cite{mason} that a suitable choice of gauge can
simplify the class of diagrams contributing at LL.
The common feature of all this work was the use of fixed order calculations.
To verify that the pattern of low order calculations survives at higher orders,
a method to demonstrate
the Regge behavior of amplitudes to all orders is necessary. This analysis was
provided  by A. Sen in Ref. \cite{sen83},
in massive QED. Sen developed a systematic way to control the high
energy behavior of fermion and photon exchange amplitudes to arbitrary
logarithmic accuracy. The formalism relies heavily
on the IR structure and gauge invariance of QED and provides a proof of the
reggeization of a fermion at NLL to all orders in
perturbation theory.

The resummation of color singlet exchange amplitudes
in non-abelian gauge theories in LL was achieved in the pioneering work
of Ref. \cite{bfkl}, where the reggeization of a gluon in LL was also
demonstrated.
The evolution equations resumming LL in the case of three gluon exchange was
derived in Ref. \cite{kwiecinski}.
In Ref. \cite{jaroszewicz}, $n$-gluon exchange amplitudes in QCD at LL level
were studied and a set of evolution
equations governing the high energy behavior of these amplitudes was
obtained at LL. A different approach was undertaken
in Ref. \cite{bartels}.
Here $n \rightarrow m$ amplitudes were studied in
SU(2) Higgs model with spontaneous symmetry breaking.
Starting with the tree level amplitudes,  an iterative
procedure was developed, which generates a minimal set of terms
in the perturbative expansion that have to be taken into account in order to
satisfy the unitarity requirement of the theory. See also Ref. \cite{cheng}.
The extension of the BFKL formalism to NLL spanned over a decade. For a
review see Ref. \cite{salam}. The building blocks
of NLL BFKL are the emissions of two gluons or two quarks along the
ladder, Ref. \cite{emission},
one loop corrections to the emission of a gluon along the ladder, Ref.
\cite{oneLoop}, and the two loop gluon trajectory,
Refs. \cite{gluonTraj}, \cite{korchemsky}, \cite{ducaGlover} and \cite{bff}.
The particular results were put together in Ref. \cite{nllBFKL}. In Ref.
\cite{ducaFadin},
the trajectory for the fermion at NLL was evaluated by taking the Regge
limit of the explicit
two loop partonic amplitudes, Ref. \cite{tejeda}.

Besides the NLO perturbative corrections to the BFKL kernel a variety of
approaches have been developed for unitarization corrections,
Refs. \cite{smallx, balitski,largeN}, which extend
the BFKL formalism by incorporating selected higher-order corrections.
The procedure proposed in this paper, in a way,
places these approaches in an even more general context.
In principle, it makes it possible to find the scattering
amplitudes to arbitrary logarithmic accuracy and to determine
the evolution kernels to arbitrary fixed order in the coupling constant.
The formalism contains all color structures and, of course, the
construction of the amplitude to any given level requires 
the computation of the kernels and the solution of 
the relevant equations.

The paper is organized as follows. In Sec. \ref{kinematics} we discuss the
kinematics of the partonic process under study
and the gauge used. In Sec. \ref{leadingRegions} we identify the leading
regions in internal momentum space,
which produce logarithmic enhancements in the perturbation series.
After identifying these regions, we perform power counting to verify that
the singularity structure
of individual diagrams is at worst logarithmic.
The leading regions lead to a factorized form for the amplitude (First
Factorized Form).
It consists of soft and jet functions, convoluted over soft loop momenta,
which can still produce logarithms of $s/|t|$.
In Sec. \ref{jet functions} we study the properties of the jet functions
appearing in the factorization formula for the amplitude.
We show how the soft gluons can be factored from the jet functions. In Sec.
\ref{evolution equations} we demonstrate how to
express systematically  the amplitude as a convolution in transverse momenta.
In this form
all the large logarithms are organized in jet functions and the soft
transverse momenta integrals do not introduce
any logarithms of $s/|t|$ (Second Factorized Form).
We derive evolution equations that enable us to control the high energy
behavior of
the scattering amplitudes. In Sec. \ref{amplitude}, we illustrate the
general methods
valid to all logarithmic accuracy in the case of LL and NLL in the amplitude
and we examine the
evolution equations at LL. Some technical details are discussed in
appendices \ref{contracted vertices} - \ref{glauberRegion}.
The first appendix treats power counting for regions of integration space where
internal loop momenta become much larger than the momentum transfer.
In Appendix \ref{variation} we illustrate the origin of special vertices
encountered in the paper.
In Appendix \ref{tulipGarden} we show a systematic expansion for the
amplitude leading to the first factorized form.
In Appendix \ref{feynmanRules} we list the Feynman rules used throughout
the text. Finally, in Appendix \ref{glauberRegion}
we demonstrate the origin of extra soft momenta configurations (Glauber
region) which need to be considered in the analysis
of amplitudes in the Regge limit.

\section{Kinematics and Gauge} \label{kinematics}

We analyze the amplitude for the elastic scattering of massless quarks
\be
q (p_A, r_A) + q'(p_B, r_B) \rightarrow q (p_A -q, r_1) + q' (p_B + q, r_2),
\label{qqqq}
\ee
within the framework of perturbative QCD in the kinematic region $s \gg -t$
({\it Regge limit}),
where $s=(p_A + p_B)^2$ and $t=q^2$ are the usual Mandelstam variables.
We stress, however, that the results obtained below apply to arbitrary
elastic two-to-two partonic process.
We pick process (\ref{qqqq}) for concreteness only.
The arguments in Eq. (\ref{qqqq}) label the momenta and the colors of the
quarks
(we do not exhibit the dependence on the polarizations).
We choose to work in the center-of-mass (c.m.) where the momenta of the
incoming quarks
and the momentum transfer have the following components
\footnote{We use light-cone coordinates,
$v=(v^+,v^-,v_{\perp})$, $v^{\pm}=(v^0 \pm v^3)/\sqrt{2}$.}
\bea \label{momentaDefinition}
p_A & = & \left(\sqrt{\frac{s}{2}}, 0^-, 0_{\perp}\right), \nonumber \\
p_B & = & \left(0^+,\sqrt{\frac{s}{2}}, 0_{\perp}\right), \nonumber \\
q   &=& (0^+, 0^-, q_{\perp}).
\eea
Strictly speaking $q^{\pm}= \pm |t| / \sqrt{2s}$, so the $q^{\pm}$
components vanish in the Regge limit only.

In the color basis
\bea \label{qqBasis}
b_{\bf 1} & = & \delta_{r_A,\,r_1} \delta_{r_B,\,r_2}, \nonumber  \\
b_{\bf 8} & = & - \frac{1}{2 N_c} \delta_{r_A,\,r_1} \delta_{r_B,\,r_2} +
\frac{1}{2} \delta_{r_A,\,r_2} \delta_{r_B,\,r_1},
\eea
with $N_c$ the number of colors, we can view the amplitude for process
(\ref{qqqq}) as a two dimensional vector in color space
\be
A = \left( \ba{c} A_{\bf 1} \\ A_{\bf 8} \ea \right),
\ee
where $A_{\bf 1}$ and $A_{\bf 8}$ are defined by the expansion
\be
A_{\, r_A \, r_B, \; r_1 \, r_2} = A_{\bf 1} \, (b_{\bf 1})_{r_A \, r_B, \;
r_1 \, r_2} + A_{\bf 8} \,
(b_{\bf 8})_{r_A \, r_B, \; r_1 \, r_2}.
\ee
Since the amplitude is dimensionless and all
particles are massless, its components can depend, in general,
on the following variables
\be
A_i \equiv A_i \left (\frac{s}{{\mu}^2},
\frac{t}{{\mu}^2},{\alpha}_s({\mu}^2),\epsilon \right) \hspace{2cm}
\mbox{for} \; i={\bf 1}, {\bf 8} ,
\ee
where $\mu$ is a scale introduced by regularization.
We use dimensional regularization in order to regulate both
infrared (IR) and ultraviolet (UV) divergences with $D = 4 - 2\varepsilon$ the
number of dimensions. Choosing the scale ${\mu}^2 = s$, the
strong coupling,
${\alpha}_s(\mu)$, is small. However, in general, an individual Feynman
diagram contributing to the process (\ref{qqqq}) at $r$-loop order can give
a contribution as singular as
$(s/t)\,{\alpha}_s^{r+1}{\ln}^{2r}(-s/t)$. In Sec.
\ref{doubleLogCancellation} we will confirm that there is a cancellation of
all terms proportional to the $i$-th logarithmic power for $i=r+1,\ldots ,
2r$ at order ${\alpha}_s^{r+1}$ in the perturbative expansion of the
amplitude. Hence at $r$ loops the amplitude is enhanced by a factor
$(s/t)\,{\alpha}_s^{r+1}\ln^r(-s/t)$, at most.
In order to get reliable results in perturbation theory we must,
nevertheless, resum these large contributions. In the $k$-th nonleading
logarithmic approximation one needs to resum all the terms proportional to $
(s/t)\,{\alpha}_s^{r+1}\ln^{r-j}(-s/t), \;\; j=0,\ldots , k $ at $r$-loop
level.

We perform our analysis in the Coulomb gauge, where the propagator of a
gluon with momentum $k$ has the form
\be \label{propagator}
-i \, {\delta}_{ab} \; \frac{N_{\alpha \beta}(k,{\bar k})}{k^2 + i\epsilon}
\equiv -i \, {\delta}_{ab} \; \frac{1}{k^2 + i\epsilon}
\left(g_{\alpha \beta} - \frac{k_{\alpha} \, {\bar k}_{\beta} + {\bar 
k}_{\alpha} \, k_{\beta}
  -k_{\alpha} \, k_{\beta}}{k \cdot {\bar k}}\right),
\ee
in terms of the vector
\be \label{barVector}
\bar k = k - (k \cdot {\eta}) \, \eta,
\ee
with
\be \label{eta}
\eta = \left(\frac{1}{\sqrt{2}},\frac{1}{\sqrt{2}},0_{\perp}\right),
\ee
an auxiliary four-vector defined in the partonic c.m. frame.
The numerator of the gluon propagator satisfies the following identities
\bea \label{gluonProperties}
k^{\alpha} \, N_{\alpha \beta}(k,{\bar k}) & = & k^2 \, \frac{k_{\beta} -
{\bar k}_{\beta}}{k \cdot {\bar k}}, \nonumber \\
{\bar k}^{\alpha} \, N_{\alpha \beta}(k,{\bar k}) & = & 0.
\eea
The first equality in Eq. (\ref{gluonProperties}) is the statement that the
nonphysical degrees of freedom do not propagate in
this gauge. For use below, we list the components of the gluon propagator:
\bea \label{propagComp}
N^{+ -} (k)  =  N^{- +}(k) & = & \frac{k^+ k^- - k_{\perp}^2}{k \cdot \bar
k}, \nonumber \\
N^{+ +} (k)  =  N^{- -}(k) & = & \frac{k^+ k^-}{k \cdot \bar k}, \nonumber
\\
N^{\pm \; i} (k) = N^{i \; \pm}(k) & = & \pm \frac{(k^- - k^+) k^i}{2 k
\cdot \bar k}, \nonumber \\
N^{i \; j} (k) = N^{j \; i}(k) & = & g^{i j} - \frac{k^i k^j}{k \cdot \bar
k}.
\eea
We note that these are symmetric functions under the transformation $k^{\pm}
\rightarrow - k^{\pm}$, except for the components
$N^{\pm \; i} = N^{i \; \pm}$, which are antisymmetric under this
transformation.
It was demonstrated in Ref. \cite{zwanziger} that QCD is renormalizable in
Coulomb gauge, by considering a class
of gauges which interpolates between the covariant (Landau) and the physical
(Coulomb) gauge.

\section{Leading Regions, Power Counting} \label{leadingRegions}

In order to resum the Regge logarithms, we need to identify the regions of
integration in the loop momentum space that give rise to singularities
in the limit $t/s\rightarrow 0$. We follow the
method developed in Refs. \cite{sterman78, powerCounting},
which begins with the identification of the relevant
regions in momentum space.

\subsection{Singular contributions and reduced diagrams}

The singular contributions of a Feynman integral come from the points in
loop momentum space where the integrand becomes singular due to the
vanishing of propagator denominators. However, in order to give a true
singularity the integration variables must be trapped at such a singular
point. Otherwise we can deform the integration contour away from the
dangerous region. These singular points are called pinch singular points.
They can be identified with the following regions of integration in momentum
space,
\begin{enumerate}
\item soft momenta, with scaling behavior $k^{\mu} \sim \sigma \, \sqrt{s}$
for all components ($\sigma \ll 1$),
\item momenta collinear to the momenta of the external particles, with
scaling behavior \\ $k^+ \sim \sqrt{s}, \; k^- \sim \lambda \, \sqrt{s}, \;
|k_{\perp}| \sim {\lambda}^{1/2} \, \sqrt{s}$ for the particles moving in
the $+$ direction and \\ $k^+ \sim \lambda \, \sqrt{s}, \; k^- \sim
\sqrt{s}, \; |k_{\perp}| \sim {\lambda}^{1/2} \, \sqrt{s}$ for the particles
moving in the $-$ direction,
\item so-called Glauber or Coulomb momenta, Ref. \cite{glauber}, with scaling
behavior $k^{\pm} \sim {\sigma}^{\pm} \, \sqrt{s}$, \, $|k_{\perp}| \sim
\sigma \, \sqrt{s}$, where $\lambda \lesssim {\sigma}^{\pm} \lesssim
\sigma$, and where the scaling factors $\lambda, \sigma$ satisfy the strong
ordering $ \lambda \ll \sigma \ll 1$ (The origin of this region is
illustrated in Appendix \ref{glauberRegion}.),
\item hard momenta, having the scaling behavior $k^{\mu} \sim \sqrt{s}$ for
all components.
\end{enumerate}
The extra gauge denominators $1/(k \cdot {\bar k})$ originating from the
numerators of the gluon propagator, Eq. (\ref{propagator}),
do not alter the classification of the pinch singular points mentioned
above. Actually, only the subsets 1 and 3 in the above classification
can be produced due to the extra gauge denominators.

\begin{figure} \center
\includegraphics*{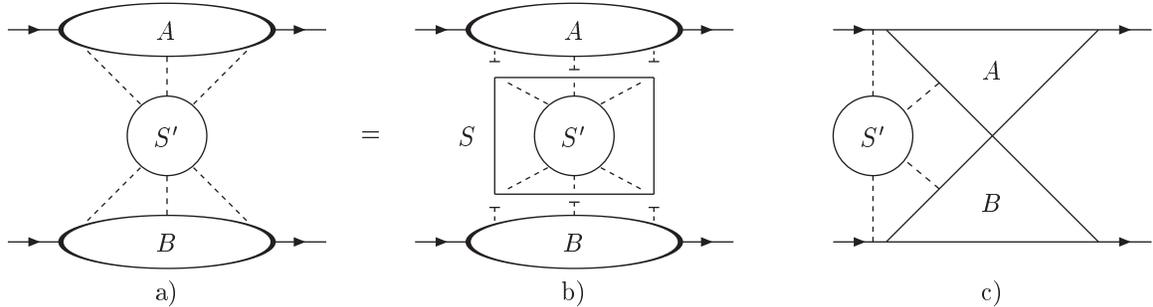}
\caption{\label{reduced} The reduced diagrams a) and c) contributing to the
amplitude. Diagram b) represents a decomposition
of diagram a) for the purpose of power counting.}
\end{figure}

With every pinch singular point, we may associate a reduced diagram, which
is obtained from the original diagram by contracting all hard lines (subset
4) at the particular singular point. As shown in Refs. \cite{sterman78, powerCounting, cono} 
the reduced diagram corresponding to a given pinch singular
point must describe a real physical process, with each vertex of the reduced
diagram representing a real space-time point. This physical interpretation
suggests two types of reduced diagrams contributing to the process
(\ref{qqqq}), shown in Fig. \ref{reduced}.

The jet $A$($B$) contains lines whose momenta represent motion in the $ + \;
(-) $ direction. The lines included in the blob $S'$
and the lines coming out of it are all soft (configurations 1 and 3 in the
classification of loop momenta described above). These two oppositely moving
(virtual) jets may interact through the exchange of soft lines, Fig.
\ref{reduced}a, and/or they can meet at one or more
space-time points, Fig. \ref{reduced}c.

Having found the most general reduced diagrams giving the leading behavior
of the amplitude for process (\ref{qqqq}) in the Regge limit, we can
estimate the strength of the IR divergence of the integral near a given
pinch singular point. First we restrict ourselves to
cases involving  subsets 1 and 2 from the classification of loop momenta above.
To do so, we count powers in the scaling variables $\lambda$
and $\sigma$.

The scaling behavior of these loop momenta implies that every soft loop
momentum contributes a factor ${\sigma}^4$, every jet loop momentum gives
rise to the power ${\lambda}^2$, every internal soft boson (fermion) line
provides a contribution ${\sigma}^{-2}$ (${\sigma}^{-1}$) and every internal
jet line (fermionic or bosonic) scales as ${\lambda}^{-1}$. In addition,
there can be suppression factors arising from the numerators of the
propagators associated with internal lines and from internal vertices.
As pointed out in Ref. \cite{sterman78}, in physical gauges each three-point
vertex connecting three jet lines is associated with a numerator factor that
vanishes  at least linearly in the components of the transverse jet momenta,
and therefore provides a suppression ${\lambda}^{1/2}$.

We are now ready to estimate the power of divergence corresponding to the
reduced diagrams describing our process. First we restrict ourselves to the
case shown in Fig. \ref{reduced}a.
As indicated schematically in Fig. \ref{reduced}b, we can perform the power
counting for the jets and for the soft part separately.
All soft propagators and all soft loop momenta are included in the soft
subdiagram $S$.
The superficial degree of IR divergence of the reduced diagram $R$ from Fig.
\ref{reduced}a and Fig. \ref{reduced}b
can then be written as
\be \label{omegaR}
\omega(R) = \omega(A) + \omega(B) + \omega(S),
\ee
where the external lines and loops of $S'$ are included in $S$.
For $\omega(R) > 0$ the overall integral is finite, while $\omega(R) \leq 0$
corresponds to an IR divergent integral.  
When $\omega(R) = 0$, the integral diverges logarithmically.
Here we set $\lambda\sim\sigma$ for power counting purposes.  
We come back to the effect of relaxing this condition 
in connection with a discussion of item 3, Glauber regions, 
in our list of singular momentum configurations.

\subsection {Power counting} \label{elementary vertices}

In this subsection, we consider the case when
all vertices in a diagram are elementary only, that is, without contracted
subdiagrams carrying large loop momenta.
In Appendix \ref{contracted vertices} we show that our conclusions are
unchanged by contracted vertices.

We perform the power counting for the soft part $S$ first. Let $f, b$ be the
number of fermion, boson lines external to $S^{'}$
and let $E=f+b$. The superficial degree of divergence for $S$, found by
summing powers of $\sigma$, can be written
\be \label{os1}
\omega(S) = 4(E-2) - 2b - f + 2 + \omega(S^{'}),
\ee
where the first term is due to loop integrations linking $S'$ to the jets,
while the second and the third terms originate from
propagators associated with the bosonic and fermionic lines, respectively,
connecting the jets $A$, $B$ and the soft part $S'$. The term $+2$ is
introduced because we are resumming only leading power corrections
proportional to $s/t$ and therefore we exclude the overall factor $s/t$ from
the power counting.
Since the lines entering $S'$ are soft, we obtain the superficial degree of
divergence for $S'$ simply from dimensional analysis.
It is given by
\be \label{osp}
\omega(S^{'}) = 4 - b - 3f/2.
\ee
Combining Eqs. (\ref{os1}) and (\ref{osp}), the superficial degree of infrared divergence 
for the soft part $S$ is then
\be \label{os}
\omega(S) = b + 3f/2 - 2.
\ee

Before carrying out the jet power counting, we introduce some notation.
Let $E_A$ be the number of soft lines attached to jet $A$; $I$ is the total
number of jet internal lines; $v_{\alpha}$ is the number of $\alpha$-point
vertices connecting jet lines only; $w_{\alpha}$ has a meaning similar to
$v_{\alpha}$, with the difference that every vertex counted by $w_{\alpha}$
has at least one soft line attached to it.
These are the vertices that connect the jet $A$ to the soft part $S$.
Finally, $L$ denotes the number of loops internal to jet $A$.
As noted above, we will perform the power counting for the case when 
the scaling factor for
the soft momenta, $\sigma$, is of the same order as the scaling factor for
jet $A$ momenta. When the scaling factors are different we encounter
subdivergencies, which can be analyzed the same way
as described below.
We also assume that there are no internal and external ghost lines included
in the jet function.
Later we will discuss the effect of adding ghost lines.

The superficial degree of divergence for jet $A$ can now be expressed as
\be \label{oa1}
\omega(A) = 2L - I + v_3/2.
\ee
The last term represents the suppression factor associated with  the three
point vertices.
We denote the total number of vertices internal to jet $A$ by
\be \label{v}
v = \sum_{\alpha}(v_{\alpha}+w_{\alpha}).
\ee
Next we use the Euler identity relating the number of loops, internal lines
and vertices of jet $A$
\be \label{euler}
L = I - v + 1,
\ee
and the relation between the number of lines and the number of vertices
\be \label{live}
2I + E_A + 2 = \sum_{\alpha} \alpha (v_{\alpha} + w_{\alpha}).
\ee

Using Eqs. (\ref{oa1})-(\ref{live}) we arrive at the following form for the
superficial degree of divergence for jet $A$
\be \label{oa2}
\omega(A) = 1 - (E_A  + w_3)/2  + \sum_{\alpha \ge 5}(\alpha - 4)(v_{\alpha}
+ w_{\alpha})/2.
\ee
Since every vertex counted by $w_{\alpha}$ connects at least one external
soft line, we have the condition
\be \label{ea}
E_A \ge w_3 + \sum_{{\alpha}\ge 4} w_{\alpha}.
\ee
The equality holds when there is no vertex with two or more soft lines
attached to it. Combining Eqs. (\ref{oa2})-(\ref{ea}) we arrive at the
following lower bound on the superficial degree of divergence for jet $A$:
\be \label{oa3}
\omega(A) \ge 1 - E_A +\sum_{{\alpha} \ge 4} w_{\alpha}/2 + \sum_{\alpha \ge
5}(\alpha - 4)(v_{\alpha} + w_{\alpha})/2.
\ee
The third and the last term in Eq. (\ref{oa3}) are always positive or zero
and hence
\be \label{oa4}
\omega(A) \ge 1 - E_A.
\ee
A similar result holds for jet $B$, and therefore the superficial degree of
collinear divergence for jets $A$ and $B$ is
\be \label{oa}
\omega(A) + \omega (B) \ge 2 - E,
\ee
with $E = E_A + E_B$ as in Eq. (\ref{os1}).
Combining the results for soft and jet power counting, Eqs. (\ref{os}) and
(\ref{oa}), respectively in Eq. (\ref{omegaR}),
we finally obtain the superficial degree of IR divergence for the
reduced diagram in Fig. \ref{reduced}a,
\be \label{or}
\omega(R) \ge f/2.
\ee
This condition says that we can have at worst logarithmic divergences,
provided no soft fermion lines are exchanged between the jets $A$ and $B$.
We can therefore conclude that a reduced diagram from Fig. \ref{reduced}a
containing elementary vertices can give at worst logarithmic enhancements in
perturbation theory. In order for the divergence to occur, the following set
of conditions must be satisfied:
\begin{enumerate}
\item There is an exchange of soft gluons between the jets $A$ and $B$ only,
with no soft fermion lines attached to the jets.
\item The jets $A$ and $B$ contain $3$ and $4$ point vertices only, see Eq.
(\ref{oa3}).
\item Soft gluons are connected to jets only through $3$ point vertices, Eq.
(\ref{oa3}), and at most one soft line is attached to each vertex inside the
jets, Eq. (\ref{ea}).
\item In the reasoning above we have assumed that there is no suppression
factor associated with the vertices where soft and jet lines meet. In order
for this to be true, the soft gluons must be connected to the jet $A$($B$)
lines via the $+(-)$ components of
the vertices.
\end{enumerate}

Next we consider adding ghost lines to the jet functions. As we review in
Appendix \ref{feynmanRules}, the propagator for a
ghost line with momentum $k$ is proportional to $1/(k \cdot {\bar k})$.
Hence every internal ghost line belonging to the jet gives a contribution
which is power suppressed as $1/s$. Since the numerator factors do not
compensate for this suppression, we can immediately conclude that the jet
functions cannot contain internal or external ghost lines at leading power.

So far we have not taken into account the possibility when the soft loop
momenta are pinched by the singularities of the jet lines. This situation
allows different components of soft momenta to scale differently. For
example, a minus component of soft momentum can scale as the minus component
of jet $A$ momentum $\lambda$, while the rest of the soft momentum
components may scale as $\sigma$, where $ \lambda \ll \sigma \ll 1$. The
origin of these extra pinches is illustrated in Appendix
\ref{glauberRegion}.

Let us see what happens when we attach the ends of a gluon line with this
extra pinch to jet $A$ at one end and the soft
subdiagram $S$ at the other end. The integration volume for this soft loop
momentum scales as $\lambda{\sigma}^3$. The soft gluon denominator gives a
factor ${\sigma}^{-2}$. If this soft gluon is connected to the soft part at
a $4$-point vertex, there is no new denominator in the soft part. On
the other hand, if the soft gluon is attached to the soft part via a
$3$-point vertex then the extra denominator including the numerator
suppression factors scales as ${\sigma}^{-1}$. The new jet line scales as
${\lambda}^{-1}$ as long as the condition ${\lambda}^{1/2} \gtrsim \sigma$
is obeyed; otherwise, we have the scaling ${\sigma}^{-2}$ for the extra jet
line. For ${\lambda}^{1/2} \gtrsim \sigma$ the Glauber region produces
logarithmic infrared divergence. When ${\lambda}^{1/2} \lesssim \sigma$, the
overall scaling factor $\lambda / {\sigma}^2$
indicates power suppressed contribution.

Let us now investigate another possibility, when the soft gluon connects jet
$A$ and jet $B$ directly and its momentum is pinched by the singularities of
the jet $A$ and the jet $B$ lines. Denoting the scaling factors of jet $A$
and jet $B$ as ${\lambda}_A$ and ${\lambda}_B$, respectively, the
integration volume provides the factor ${\lambda}_A
{\lambda}_B {\sigma}^2$ and the soft gluon denominator contributes the power
${\sigma}^{-2}$. The extra jet $A$ and jet $B$ denominators scale as
${\lambda}_A^{-1}$ and
${\lambda}_B^{-1}$, provided ${\lambda}_A^{1/2} \gtrsim \sigma$ and
${\lambda}_B^{1/2} \gtrsim \sigma$. For ${\lambda}_{A,B}^{1/2} \lesssim
\sigma$ both extra jet denominators provide the scaling factor
${\sigma}^{-2}$. When ${\lambda}_{A, B}^{1/2} \gtrsim \sigma$, the
power counting suggests logarithmically divergent integrals.

We have therefore verified that when the softest component of a soft line
satisfies the ordering
${\sigma}^2 \lesssim \lambda \lesssim \sigma$, the Glauber (Coulomb) momenta
produce logarithmically IR divergent integrals and need to
be taken into an account when identifying enhancements in perturbation
series.  The analysis demonstrated above for the case of one Glauber 
gluon can be
extended to the situation with arbitrary
number of Glauber gluons. This follows from dimensional
analysis, in a similar fashion as the treatment of purely soft loop momenta above.

We conclude that the reduced diagram in Fig. \ref{reduced}a is at most
logarithmically IR divergent, modulo the factor $s/|t|$. The reduced diagram
in Fig. \ref{reduced}b looses one small denominator compared to the reduced
diagram in Fig. \ref{reduced}a and since we are working in physical gauge,
this loss cannot be compensated by a large kinematical factor coming from
the numerator. Hence the reduced diagram in Fig. \ref{reduced}b is power
suppressed compared to the reduced diagram in Fig. \ref{reduced}a, and we do
not need to consider it at leading power.

Finally, let us discuss the scale of the soft momenta. In the case of soft
exchange lines, each gluon propagator supplies a factor $1/({\sigma}^2 \,
{s})$, which we want to keep at or below the order $t$ in the leading power
approximation.   Thus the size of the scale is fixed to be $\sigma \sim
\sqrt{|t|}/\sqrt{s}$. In the case of soft lines which are attached to jet
$A$ or to jet $B$ only, the scaling factor lies in the interval
$(\sqrt{|t|}/\sqrt{s}, 1)$. In the case of Glauber momenta, we again
need $\sigma \sim  \sqrt{|t|}/\sqrt{s}$. Then the condition
${\lambda}^{1/2} \gtrsim \sigma$, which is necessary for the logarithmic
enhancement, implies that the scaling factors for $+$ and $-$
components of the Glauber (Coulomb) momenta can go down to $|t|/s$, the scale
of the small components of jet momenta.
Additionally, we should note that soft and jet subdiagrams that do 
not carry the momentum
transfer may approach the mass shell ($\lambda,\ \sigma\rightarrow 0$).
Such lines produce true infrared divergences, which we assume
are made finite by dimensional regularization to preserve the gauge
properties that we will use below.  The same power counting as above shows that
these divergences are also at worst logarithmic.

\begin{figure} \center
\includegraphics*{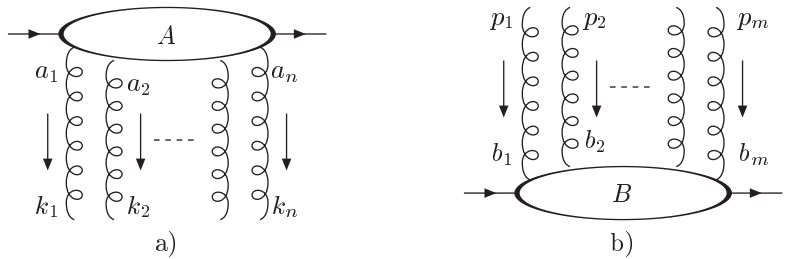}
\caption{\label{jetA} Jet $A$ moving in the $+$ direction (a) and jet $B$ moving in the $-$ direction (b).}
\end{figure}

\subsection{First factorized form} \label{first factorized form}

The analysis of the previous subsection suggests the following decomposition
of the leading reduced diagram from Fig. \ref{reduced}a. Let us denote the
$(n+2)$-point and $(m+2)$-point  Green
functions, 1PI in external soft gluon lines, corresponding to jet 
$A$, $J_{(A) \, {\mu}_1 \ldots \,
{\mu}_n}^{(n) \, a_1 \ldots \, a_n} (p_A, q, \eta; k_1, \ldots, k_n)$, Fig.
\ref{jetA}a, and to jet $B$, $J_{(B) \, {\nu}_1 \ldots \, {\nu}_m}^{(m) \,
b_1 \ldots \, b_m}(p_B, q, \eta; p_1, \ldots, p_m)$, Fig. \ref{jetA}b,
respectively. The jet function $J_{(A)}^{(n)}$ ($J_{(B)}^{(m)}$) also
depends on the color of the incoming and outgoing partons $r_A$, $r_1$
($r_B$, $r_2$), as well as on their
polarizations ${\lambda}_A$, ${\lambda}_1$ (${\lambda}_B$, ${\lambda}_2$),
respectively. In order to avoid making the notation even more cumbersome we do
not exhibit this dependence explicitly. In addition the dependence of
$J_{(A)}^{(n)}$ and $J_{(B)}^{(m)}$ on the renormalization scale $\mu$ and
the running coupling ${\alpha}_s(\mu)$ is understood. The jet functions also
depend on the following parameters: the gauge fixing vector $\eta$, Eq. (\ref{eta}), of the
Coulomb gauge, the four momenta of the external soft gluons attached to jet
$A$ ($B$), $k_1, \ldots, k_n$ ($p_1, \ldots, p_m$), and the Lorentz and
color indices of the soft gluons attached to the jet $A$ ($B$), ${\mu}_1,
\ldots, {\mu}_n$; $a_1, \ldots, a_n$ (${\nu}_1, \ldots, {\nu}_m$; $b_1,
\ldots, b_m$). The momenta of the soft gluons attached to the jets $A$ and
$B$ satisfy the constraints $\sum_{i=1}^{n} k_i = q$ and
$\sum_{j=1}^{m} p_j = q$.

According to the results of the power counting, the soft gluons couple to
jet $A$ via the minus components of their polarizations, and
to jet $B$ via the plus components of their polarizations.
Therefore, only the following components survive in the leading power
approximation
\bea \label{jetab}
J_{A}^{(n) \, a_1 \ldots \, a_n} (p_A, q, \eta, v_B; k_1, \ldots, k_n) &
\equiv & \left(\prod^n_{i=1} v_B^{{\mu}_i} \right) J^{(n) \, a_1 \ldots \,
a_n}_{(A) \, {\mu}_1 \ldots \, {\mu}_n}(p_A, q, \eta; k_1, \ldots, k_n),
\nonumber \\
J_{B}^{(m) \, b_1 \ldots \, b_m} (p_B, q, \eta, v_A; p_1, \ldots, p_m) &
\equiv & \left(\prod^m_{i=1} v_A^{{\nu}_i} \right) J^{(m) \, b_1 \ldots \,
b_m}_{(B) \, {\nu}_1 \ldots \, {\nu}_m}(p_B, q, \eta; p_1, \ldots, p_m),
\eea
where we have defined light-like momenta in the plus direction $v_A = (1, 0,
0_{\perp})$ and in the minus direction $v_B = (0, 1, 0_{\perp})$. We can now
write the contribution to the reduced diagram in Fig. \ref{reduced}a, and
hence to the amplitude for process (\ref{qqqq}), in the form
\bea \label{fact1}
A & = & \sum_{n, m} \int \left(\prod^{n-1}_{i=1} \mathrm{d}^D k_i \right) \,
\int \left(\prod^{m-1}_{j=1} \mathrm{d}^D p_j \right) \, J_A^{(n) \, a_1
\ldots \, a_n} (p_A, q, \eta, v_B; k_1, \ldots, k_n) \nonumber \\
& \times & S^{(n, m)}_{a_1 \ldots \, a_n, b_1 \ldots \, b_m} (q, \eta, v_A,
v_B; k_1, \ldots, k_n; p_1, \ldots, p_m) \nonumber \\
& \times & J_B^{(m) \, b_1 \ldots \, b_m} (p_B, q, \eta, v_A; p_1, \ldots,
p_m),
\eea
where the sum over repeated color indices is understood.
Corrections to Eq. (\ref{fact1}) are suppressed by positive powers of
$t/s$.
The jet functions $J_{A,B}$ are defined in Eq. (\ref{jetab}) in the leading
power accuracy.
The internal loop momenta of the jets $A$, $B$ and of the soft
function $S$ are integrated over. The soft function will, in general,
include delta functions setting some of the momenta $k_1, \ldots \, , k_n$
and color indices $a_1, \ldots \, , a_n$ of jet function $J_A$ to the
momenta $p_1, \ldots \, , p_m$ and to the color indices $b_1, \ldots \, ,
b_m$ of jet function $J_B$. The construction of the soft function $S$ is
described in Appendix \ref{tulipGarden}. For a given Feynman diagram there
exist many reduced diagrams of the type shown in Fig. \ref{reduced}a, and one
has to be careful in systematically expanding this diagram into the terms
that have the form of Eq. (\ref{fact1}). This systematic method can be
achieved using the ``tulip-garden'' formalism first introduced in Ref. \cite{coso} and
used in a similar context in Ref. \cite{sen83}. For convenience of the
reader we summarize this procedure in
Appendix \ref{tulipGarden}.

Let us now identify the potential sources of the enhancements
in $\ln(s/|t|)$ of the amplitude given by Eq. (\ref{fact1}). If we 
integrate over the
internal momenta of the jet functions then we can get $\ln((p_A \cdot
\eta)^2 / |t|)$ from $J_A$ and $\ln((p_B \cdot \eta)^2/|t|)$ from $J_B$. In
addition, according to the results of the power-counting, Eq. (\ref{oa4}),
we know that the jet function with $n$ external soft gluons diverges as
$1/{\lambda}^{n-1}$. After performing the integrals over the minus
components of the external soft gluon lines attached to jet $A$ and over the
plus components of the external soft gluons connected to jet $B$, these
divergent factors are
potentially converted into logarithms of $\ln((p_A \cdot \eta)^2/|t|)$
and $\ln((p_B \cdot \eta)^2/|t|)$, respectively. Our goal will be to separate
the full amplitude into a convolution over parameters that do not introduce
any further logarithms of the form $\ln(s/|t|)$. This task will be achieved
in Sec. \ref{secondFactorizedForm}. In the following section,
we analyze the characteristics of the jet functions.

\section{The Jet Functions} \label{jet functions}

In this section we study the properties of the jet functions $A$, $B$ given
by Eq. (\ref{jetab}) since, as Eq. (\ref{fact1}) suggests, they will play an
essential role in later analysis. Since the methods for both jet functions
are similar we restrict our analysis to jet $A$ only; jet $B$ can be worked
out in the same way.
In Sec. \ref{jet1} we examine the properties of jet $A$ when the minus
component of one of its external soft gluon momenta is of order
$\sqrt{|t|}$. In Sec. \ref{jet2} we find the variation of jet $A$ with
respect to the gauge fixing vector $\eta$, and finally in Sec. \ref{jet3} we
examine the dependence of jet $A$ on the plus component of a soft gluon
momentum attached to this jet.

\subsection{Decoupling of a soft gluon from a jet} \label{jet1}

According to the results of power counting above, soft gluons attach 
to lines in
jet $A$ via the minus components of their polarization. Following the
technique of Grammer and Yennie \cite{graye} we decompose the vertex at
which the $j$th gluon is connected to jet A.
We start with a trivial rewriting of $J_A$ in Eq. (\ref{jetab})
\begin{figure} \center
\includegraphics*{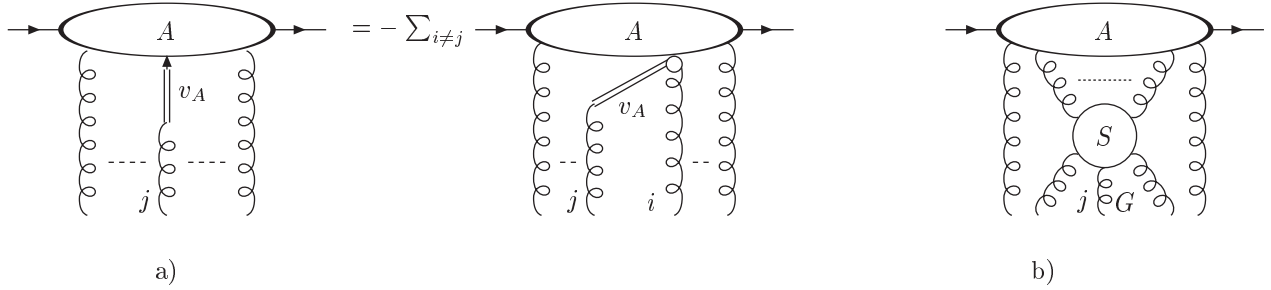}
\caption{\label{decoupling} a) Decoupling of a $K$ gluon from jet $A$.
b) Leading contributions resulting from the attachment of a $G$ gluon to
jet $A$.}
\end{figure}
\be \label{gy1}
J_A^{(n) \, a_1 \ldots \, a_n} = \left( \prod^{n}_{i \ne j} v_B^{{\mu}_i}
\right) \,
v_B^{{\mu}_j} \, g_{{\mu}_j}^{\;\;\; {\nu}_j} \, J_{(A) \; {\mu}_1 \ldots \,
{{\nu}_j} \ldots \, {{\mu}_n}}^{(n) \, a_1 \ldots \, a_n}.
\ee
We now decompose the metric tensor into the form $g^{\mu\nu} = K^{\mu\nu}(k_j)
+ G^{\mu\nu}(k_j)$ where for a gluon with
momentum $k_j$ attached to jet $A$, $K^{\mu\nu}$ and $G^{\mu\nu}$ are 
defined by
\bea \label{kgDef}
K^{\mu\nu}(k_j) & \equiv & \frac{v_A^{\mu} \, k_j^{\nu}}{v_A \cdot k_j -
i\epsilon} \nonumber \\
G^{\mu\nu}(k_j) & \equiv & g^{\mu\nu} - K^{\mu\nu} (k_j) \label{gy2}.
\eea
The $K$ gluon carries scalar polarization. Since the jet $A$ function has no
internal tulip-garden subtractions (they are contained in the soft function
$S$), we can use the Ward identities of the theory \cite{thooft},
which are readily derived  from its underlying BRS symmetry
\cite{brs}, to decouple this gluon from the rest of the jet $A$ after we sum
over all possible insertions of the gluon. The result is
\bea \label{gy2}
J_A^{(n) \, a_1 \ldots \, a_j \ldots \, a_n} (p_A, q, v_B, \eta; k_1,
\ldots, k_i, \ldots, k_j, \ldots, k_n)
& = & - \frac{1}{v_A \cdot k_j - i\epsilon} \sum^n_{i \neq j} \left(- ig_s
f^{c_i a_i a_j}\right) \nonumber \\
& & \hspace*{-5cm} \times \, J_A^{(n-1) \, a_1 \ldots \, c_i \ldots \,
{\underline a}_j \, \ldots \, a_n}
(p_A, q, v_B, \eta; k_1, \ldots, k_i + k_j, \ldots, {\underline k}_j, \ldots
k_n). \nonumber \\
\eea
The notation ${\underline a}_j$ and ${\underline k}_j$ indicates that the
jet function $J_A^{(n-1)}$ does not depend on the color index $a_j$ and the
momentum $k_j$, because they have been factored out. In Eq. (\ref{gy2}),
$g_s$ is the QCD coupling constant and $f^{c_i a_i a_j}$ are the structure
constants of the $SU(3)$ algebra. The pictorial representation of this
equation is shown in Fig. \ref{decoupling}a. The arrow represents a scalar
polarization and the double line stands for the eikonal line. The Feynman
rules for the special vertices
and the eikonal lines in Fig. \ref{decoupling}a are listed in Appendix
\ref{feynmanRules}. Strictly speaking the right-hand side
of Eq. (\ref{gy2}) and Fig. \ref{decoupling}a contain contributions
involving external ghost lines.
However, from the power counting arguments of Sec. \ref{elementary vertices}
we know that when all lines inside of the jet are jet-like, the jet function
can contain neither external nor
internal ghost lines. Therefore Eq. (\ref{gy2}) is valid up to power
suppressed corrections for this momentum configuration.

The idea behind the $K$-$G$ decomposition is that the contribution of the
soft $G$ gluon attached to the jet line in the leading power is proportional
to $v_B^{\mu} G_{\mu\nu} v_A^{\nu} = 0$.
In order to avoid this suppression, the $G$ gluon must be attached to a soft
line. The general reduced diagram corresponding to the $G$ gluon attached to
jet $A$ is depicted in Fig. \ref{decoupling}b.
The lines coming out of $S$ as well as the lines included in it are soft.
The letter $G$ next to the $j$th gluon in Fig. \ref{decoupling}b reminds us
that this gluon is a $G$-gluon attaching to jet
$J_{(A) \, \mu}$ via the $G^{+\mu}(k_j)$ vertex.

The reasoning described above applies to the case when all components of
soft momenta are of the same order. In the situation of Coulomb (Glauber)
momenta, this picture is not valid anymore, since the large ratio
$k_{\perp}/k^-$ coming from the $G^{+\perp}$ component can compensate for
the suppression due to the attachment of the $G$ part to a jet $A$ line via
the transverse components of the vertex.

\subsection{Variation of a jet function with respect to a gauge fixing vector $\eta$} \label{jet2}

In this subsection we find the variation of the jet function $J_A^{(n)}$ 
with respect to a gauge fixing vector $\eta$. 
The motivation to do this can be easily understood. 
We consider the jet function with one soft gluon attached to it only, $J_A^{(1)}(p_A, q, v_B, \eta)$. 
Let us define
\be \label{xidef}
{\xi}_A \equiv p_A \cdot \eta \;\; \mbox{and} \;\; {\zeta}_B \equiv \eta \cdot v_B.
\ee
In these terms, jet function $J_A^{(1)}$ can depend on the following kinematical combinations: \\
$J_A^{(1)}(p_A, q, v_B, \eta) = J_A^{(1)}({\xi}_A, \, p_A \cdot v_B, \, {\zeta}_B, \, t)$. Using the identity 
$p_A \cdot v_B = 2 \, {\xi}_A {\zeta}_B$ and the fact, 
that the dependence of $J_A$ on the vector $v_B$ is introduced trivially via Eq. (\ref{jetab}), we conclude that 
\be \label{jetDependence}
J_A^{(1)}(p_A, q, v_B, \eta) = {\zeta}_B \, {\bar J}_A^{(1)}({\xi}_A, \, t).
\ee
Our aim is to resum the large logarithms of $\ln (p_A^+)$ that appear in the perturbative expansion of the jet $A$ function. 
In order to do so, we shall derive an evolution equation for $p_A^+ \, \partial J^{(1)}_A / \partial p_A^+$. 
Since $p_A$ appears in combination with $\eta$ only, we can trace out the $p_A^+$ dependence of $J_A^{(1)}$ 
by tracing out its dependence on $\eta$. This can be achieved by varying the gauge fixing vector $\eta$. 
The idea goes back to Collins and Soper \cite{coso} and Sen \cite{sen81}. We will generalize the result to $J_A^{(n)}$ 
in Sec. \ref{evoleq}. 

We consider a variation that corresponds to an infinitesimal Lorentz boost in a positive $+$ direction with velocity $\delta \beta$. 
Thus, for the gauge fixing vector $\eta =(1,0,0,0)$ \footnote{For the moment we use Cartesian coordinates.}, Eq. (\ref{eta}), 
the variation is: $\delta \eta \equiv {\tilde \eta} \, \delta \beta \equiv 
(0,0,0,1) \, \delta \beta$. It leaves invariant the norm ${\eta}^2 = 1$ to order ${\cal O}(\delta \beta)$.
The precise relation between the variation of the jet $A$ function with respect to $p_A^+$ and $\delta {\eta}^{\alpha}$ is
\be \label{evolut}
p_A^+ \frac{\partial J_A^{(1)}}{\partial \, p_A^+} = - {\tilde \eta}^{\alpha}\frac{\partial J_A^{(1)}}{\partial 
\, \eta^{\alpha}} + {\zeta}_B\frac{\partial J_A^{(1)}}{\partial {\zeta}_B} = - 
{\tilde \eta}^{\alpha}\frac{\partial J_A^{(1)}}{\partial \, \eta^{\alpha}} + J_A^{(1)}. 
\ee
We have used the chain rule in the first equality and the simple relation 
${\zeta}_B \, \partial J_A^{(1)}/ \partial {\zeta}_B = J_A^{(1)}$, following from Eq. (\ref{jetDependence}), in the second one.
 
\begin{figure} \center 
\includegraphics*{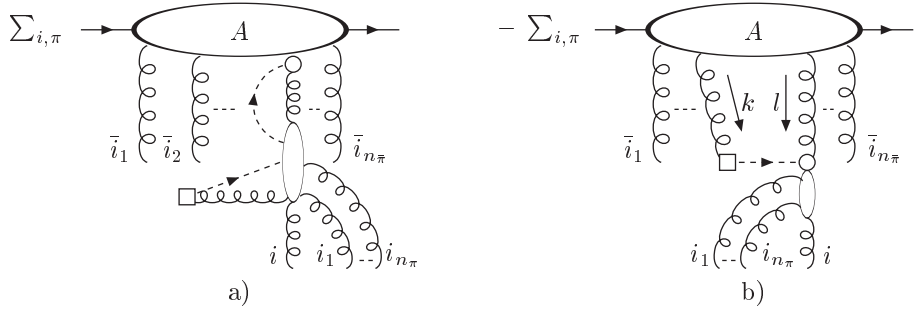}
\caption{\label{variationJ} The result of a variation of jet function $J_A^{(n)}$ with respect to a gauge fixing vector.} 
\end{figure}
 
In order for Eq. (\ref{evolut}) to be useful, we need to know what the variation of jet $A$ with respect to the gauge fixing 
vector $\eta$ is. The result of this variation for $J_A^{(n)}$ is shown in Fig. \ref{variationJ}. 
It can be derived using either the formalism of the 
effective action, Ref. \cite{nielsen}, or a diagrammatic approach first suggested in Ref. \cite{coso} and performed in axial gauge. 
We give an argument how Fig. \ref{variationJ} arises in Appendix \ref{variation}. 
Here we only note that the form of the diagrams in Fig. \ref{variationJ}
is a direct consequence of a 1PI nature of the jet functions. The explicit form of the boxed vertex 
\be \label{boxedVertex}
-i \, S^{\alpha}(k) \equiv -i \left( \eta \cdot k \, {\tilde \eta}^{\alpha} + {\tilde \eta} \cdot k \, \eta^{\alpha} \right),
\ee
as well as of the circled vertex is given in Fig. \ref{feynmanRulesF} of Appendix \ref{feynmanRules}, 
while their origin is demonstrated in Appendix \ref{variation}.
The dashed lines in Fig. \ref{variationJ} represent ghosts, and these are also given in Fig. \ref{feynmanRulesF} 
of Appendix \ref{feynmanRules}.
The four vectors $\eta$, given in Eq. (\ref{eta}), and 
\bea \label{etaTilde}
{\tilde \eta} & = & \left(\frac{1}{\sqrt{2}},- \frac{1}{\sqrt{2}},0_{\perp}\right),
\eea
appearing in Eq. (\ref{boxedVertex}) are defined in the partonic c.m. frame, Eq. (\ref{momentaDefinition}). 
We list the components of $S_{\mu} \, N^{\mu \, \alpha}(k)$
\bea \label{sCompon}
S_{\mu}(k) \, N^{\mu \, \pm}(k) & = & k^{\mp} \left( \frac{k_+^2 - k_-^2}{2 \, k \cdot \bar k} \pm 1 \right), \nonumber \\
S_{\mu}(k) \, N^{\mu \, i}(k) & = & \frac{k_-^2 - k_+^2}{2 k \cdot \bar k} \, k^i, 
\eea
for later reference.

In Fig. \ref{variationJ}, we sum over all external gluons. This is 
indicated by the sum over $i$.
In addition, we sum over all possible 
insertions of external soft gluons 
$\{i_1, \ldots, i_{n_{\pi}}\} \in 
\{1, \ldots, n\} \backslash \{i\}$. 
This summation is denoted by the 
symbol $\pi$.
We note that at lowest order, with only a gluon $i$ 
attached to the vertical blob in 
Fig. \ref{variationJ}b, this 
vertical blob denotes
the transverse tensor structure depending on the momentum $k_i$
of this gluon
\be \label{invPropag}
i\left(k_i^2 g^{\alpha\beta} - k_i^{\alpha} 
k_i^{\beta}\right).
\ee
It is labeled by a gluon line which is 
crossed by two vertical lines, Fig. \ref{feynmanRulesF}. 
The ghost 
line connecting the boxed and the circled vertices in Fig. 
\ref{variationJ}b can interact with jet $A$ 
via the exchange of an 
arbitrary number of soft gluons.
We do not show this possibility in 
Fig. \ref{variationJ}b for brevity.  

Let us now examine what the 
important integration regions for a loop with momentum $k$ in Fig. 
\ref{variationJ}b are. 
The presence of the ghost line and of the 
nonlocal boxed vertex requires that in the leading power
the loop momentum $k$ must be soft. It can be neither collinear nor 
hard. This will enable us to
factor the gluon with momentum $k$ from the rest of the jet according 
to the procedure described in
Sec. \ref{jet1}.

\subsection{Dependence of a jet function on the plus component of a 
soft gluon's momentum attached to it} \label{jet3}

\begin{figure} \center
\includegraphics*{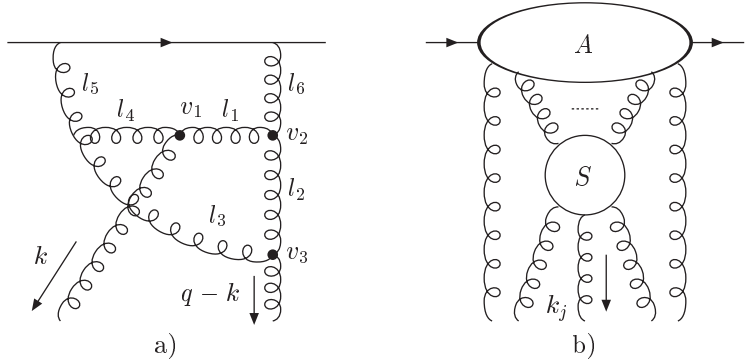}
\caption{\label{plusVar} a) Momentum flow of the external soft gluon 
inside of jet $A$.
b) Typical contribution to $k_j^+ {\partial J_A^{(n)}} / {\partial} k_j^+$.}
\end{figure}

In this subsection we want to find the leading regions of the object 
$k_j^+ {\partial J_A^{(n)}} / {\partial} k_j^+ $.
This information will be essential for the analysis pursued in the 
next sections. For a given diagram contributing to $J_A^{(n)}$ we can
always label the internal loop momenta in such a way that the 
momentum $k_j$ flows along a continuous path connecting the vertices 
where the
momentum $k_j$ enters and leaves the jet function $J_A^{(n)}$. When 
we apply the operation $k_j^+ \partial / \partial k_j^+$ on a 
particular
graph corresponding to $J_A^{(n)}$, it only acts on the lines and 
vertices which form this path.  The idea is illustrated in Fig.
\ref{plusVar}a. The gluon with momentum $k$ attaches to jet $A$ via 
the three-point vertex $v_1$. Then the momentum $k$ flows through the
path containing the vertices $v_1, v_2, v_3$ and the lines $l_1, 
l_2$. The action of the operator $k^+ \partial / \partial k^+$ on a 
line or
vertex which carries jet-like momentum  gives a negligible 
contribution, since the $+$ component of this lines momentum will be 
insensitive
to $k^+$. In order to get a non-negligible contribution, the 
corresponding line must be soft. In Fig.\ \ref{plusVar}a, lines $l_1$ 
and $l_2$
must be soft in order to get a non-suppressed contribution from the 
diagram after we apply the $k^+ \partial / \partial k^+$ operation on 
it.
This, with the fact that the external soft gluons carry soft momenta, 
also implies that the lines $l_3,
\ldots, \, l_6$ must be soft. This reasoning suggests that in general 
a typical contribution to $k_j^+ {\partial
J_A^{(n)}} / {\partial} k_j^+$ comes from the configurations shown in 
Fig. \ref{plusVar}b. It can be represented as
\bea \label{varPlus1}
J_A^{(n) \, a_1 \, \ldots \, a_n} & = & \int \left( 
\prod_{i=1}^{n'-1} {\mathrm d}^D k'_i \right) \, \,  j^{(n, n') \, 
a_1 \, \ldots
\, a_n, \, a'_1 \, \ldots \, a'_{n'}} (v_A, q, \eta; k_1, \ldots, 
k_n; k'_1, \ldots, k'_{n'}) \nonumber \\
  & \times & J_A^{(n') \, a'_1 \, \ldots \, a'_{n'}} (p_A, q, \eta, 
v_B; k'_1, \ldots, k'_{n'}).
\eea
The function $j^{(n, n')}$ contains the contributions from the soft 
part $S$ and from the gluons
connecting the jet $J_A^{(n')}$ and $S$ in Fig. \ref{plusVar}b. The 
jet function $J_A^{(n')}$ has fewer  loops
than the original jet function $J_A^{(n)}$. Now applying the 
operation $k_j^+ {\partial} / {\partial} k_j^+ $ to Eq.
(\ref{varPlus1}), the operator $k_j^+ {\partial} / {\partial} k_j^+ $ 
acts only to the function $j^{(n, n')}$. Hence we
can write
\bea \label{varPlus2}
k_j^+ \frac{\partial}{k_j^+} J_A^{(n) \, a_1 \, \ldots \, a_n} & = &
  \int \left( \prod_{i=1}^{n'-1} {\mathrm d}^D k'_i \right) 
k_j^+\frac{\partial}{\partial k_j^+} \, j^{(n, n') \, a_1 \, \ldots \,
a_n, \, a'_1 \, \ldots \, a'_{n'}} (v_A, q, \eta; k_1, \ldots, k_n; 
k'_1, \ldots, k'_{n'}) \nonumber \\
& \times & J_A^{(n') \, a'_1 \, \ldots \, a'_{n'}} (p_A, q, \eta, 
v_B; k'_1, \ldots, k'_{n'}).
\eea
We conclude that the contribution to $k_j^+ {\partial J_A^{(n)}} / 
{\partial} k_j^+$ can be expressed in
terms of jet functions
$J_A^{(n')}$ which have fewer loops than the original jet function.

\section{Factorization and Evolution Equations} \label{evolution equations}

We are now ready to obtain evolution equations which will enable us 
to resum the large logarithms.
First, in Sec. \ref{secondFactorizedForm}, we will put Eq. 
(\ref{fact1}) into what we call the second factorized form.
Then,  in Sec. \ref{evoleq}, we derive the desired evolution 
equations. In Sec. \ref{doubleLogCancellation}, we will show the cancellation of 
the double logarithms and finally in Sec. \ref{solEvolEq}, we demonstrate that the evolution equations
derived in Sec. \ref{evoleq} are sufficient to determine the high-energy behavior of the scattering 
amplitude.

\subsection{Second factorized form} \label{secondFactorizedForm}

The goal of this subsection is to rewrite Eq. (\ref{fact1}) into the 
following form \cite{sen83}
\bea \label{fact2}
A & = & \sum_{n,m} \int \left( \prod_{i=1}^{n-1} \mathrm{d}^{D-2} 
k_{i \perp} \right)
\left( \prod_{j=1}^{m-1} \mathrm{d}^{D-2} p_{j \perp} \right) \, 
{\Gamma}_A^{(n) \, a_1 \ldots \, a_n}(p_A, q, \eta, v_B; k_{1 \perp}, 
\ldots, k_{n \perp}; M) \nonumber \\
& \times & S^{' \, (n,m)}_{a_1 \ldots \, a_n, \, b_1 \ldots \, 
b_m}(q, \eta, v_A, v_B; k_{1 \perp}, \ldots, k_{n \perp}; p_{1 
\perp}, \ldots, p_{m \perp}; M) \nonumber \\
& \times & {\Gamma}_B^{(m) \, b_1 \ldots \, b_m}(p_B, q, \eta, v_A; 
p_{1 \perp}, \ldots, p_{m \perp}; M),
\eea
where ${\Gamma}_A^{(n)}$ and ${\Gamma}_B^{(m)}$ are defined as the 
integrals of the jet
functions $J_A^{(n)}$ and $J_B^{(m)}$, over the minus
and plus components, respectively, of their external soft momenta, 
with the remaining light-cone
components of soft momenta set to zero,
\bea \label{gammaDef}
{\Gamma}_A^{(n) \, a_1 \ldots \, a_n} (p_A, q, \eta, v_B; k_{1 
\perp}, \ldots, k_{n \perp}; M) & \equiv & \prod_{i=1}^{n-1} \left( 
\int_{-M}^{M} \mathrm{d} k^-_i \right ) \, J_A^{(n) \, a_1 \ldots \, 
a_n}(p_A, q, \eta, v_B; k_{1\perp}, \ldots, k_{n\perp}, \nonumber \\
& & k_1^+ = 0, \ldots, k_n^+ = 0, k_1^-, \ldots, k_n^-), \nonumber \\
{\Gamma}_B^{(m) \, b_1 \ldots \, b_m} (p_B, q, \eta, v_A; p_{1 
\perp}, \ldots, p_{m \perp}; M) & \equiv & \prod_{i=1}^{m-1} \left( 
\int_{-M}^{M} \mathrm{d} p^+_i \right ) \, J_B^{(m) \, b_1 \ldots \, 
b_m}(p_B, q, \eta, v_A; p_{1\perp}, \ldots, p_{m\perp}, \nonumber \\
& & p_1^- = 0, \ldots, p_m^- = 0, p_1^+, \ldots, p_m^+).
\eea
In Eq. (\ref{fact2}), $S'$ is a calculable function of its arguments 
and $M$ is an arbitrary scale of the order $\sqrt{|t|}$. The 
functions ${\Gamma}_{A,B}$ and $S'$ depend individually on this 
scale, but the final result, of course, does not. Based on the 
discussion at the end of Sec. \ref{first factorized form}, one can 
immediately recognize that all the large logarithms are now contained 
in the functions ${\Gamma}_A$ and ${\Gamma}_B$. The convolution of 
${\Gamma}_A, {\Gamma}_B$ and $S'$ is over the transverse momenta of 
the exchanged soft gluons. Since these momenta are restricted to be 
of the order $\sqrt{|t|}$, the integration over transverse momenta 
cannot introduce $\ln(s/|t|)$.
This indicates that at leading logarithm approximation the 
factorized diagram with the exchange of one gluon only contributes.
In general, when we consider a contribution to the amplitude at $ L 
= L_A + L_B+L_{S'}$ loop level, where $L_A, L_B$ and $L_{S'}$ is the 
number of loops in ${\Gamma}_A, {\Gamma}_B$ and $S'$,
respectively, we can get $L - L_{S'}$ logarithms of $s/|t|$ at most.
Hence, the investigation of the $s/t$ dependence of the full 
amplitude reduces to the study of the $p_A^+$ and $p_B^-$ dependence 
of ${\Gamma}_A$ and ${\Gamma}_B$, respectively. We formalize this 
statement at the end of Sec. \ref{doubleLogCancellation} after we 
have proved that $\Gamma_A$ ($\Gamma_B$) contains one logarithm of 
$p_A^+$ ($p_B^-$) per loop.

Let us now show how we can systematically go from Eq. (\ref{fact1}) 
to Eq. (\ref{fact2}). We follow the method
developed in
Ref. \cite{sen83}.
We start from Eq. (\ref{fact1}) and consider the $k_i^-$ integrals 
over the jet function $J_A$ for fixed $k^+_i, \, k_{i \perp}$:
\be \label{minusint}
A = \sum_n \int \prod_{i=1}^{n-1} {\mathrm d}k_i^- \; R_A^{\, a_1 
\ldots \, a_n}(k_1^-, \ldots, k_n^-; \ldots) \, J_A^{(n) \, a_1 
\ldots \, a_n}(p_A, q, \eta, v_B; k_1, \ldots, k_n),
\ee
where $R_A$ is given by the soft function $S$ and the jet function $J_B$,
\bea \label{ra}
R_A^{\, a_1 \ldots \, a_n}(k_1^-, \ldots, k_n^-; \ldots) & = & 
\sum_{m} \int \left(\prod^{m-1}_{j=1} \mathrm{d}^D p_j \right) \,
S^{(n, m)}_{a_1 \ldots \, a_n, b_1 \ldots \, b_m} (q, \eta, v_A, v_B; 
k_1, \ldots, k_n; p_1, \ldots, p_m) \nonumber \\
& \times & J_B^{(m) \, b_1 \ldots \, b_m} (p_B, q, \eta, v_A; p_1, 
\ldots, p_m).
\eea
We next use the following identity for $R_A$: \footnote{Recall that 
$k_n = q - (k_1 + \ldots \, +k_{n-1})$, so $k_n$ is
not an independent momentum.}
\bea \label{iden}
R_A(k_1^-, \ldots, k_{n-1}^-) & = & R_A(k_1^-=0, \ldots, k_{n-1}^-=0) 
\, \prod_{i=1}^{n-1} {\theta}(M-|k_i^-|) \nonumber \\
& + &
\sum^{n-1}_{i=1} \; \left[ R_A(k_1^-, \ldots, k_i^-, k_{i+1}^-=0, 
\ldots, k_{n-1}^- =0) \right.
\nonumber \\
& - & \left.
R_A (k_1^-, \ldots, k_{i-1}^-, k_{i}^-=0, \ldots, k_{n-1}^- =0) \, 
{\theta}(M-|k_i^-|) \right] \nonumber \\
& \times & \prod_{j=i+1}^{n-1} {\theta} (M-|k_j^-|).
\eea
We have suppressed the dependence on the color indices and other 
possible arguments in $R_A$ for brevity. The scale $M$ can be 
arbitrary, but, as above, we take it to be of the order of 
$\sqrt{|t|}$. The first term on the right hand side of Eq. 
(\ref{iden}) has all $k_i^- = 0$. The rest of the terms can be 
analyzed using the $K$-$G$ decomposition discussed in Sec. 
\ref{jet1}. Consider the ($i=1$) term, say, in the square bracket of 
Eq. (\ref{iden}) inserted in Eq. (\ref{minusint}). Let us denote it 
$A_1$. In the region $|k_1^-| \ll M$ the integrand vanishes. On the 
other hand, for $|k_1^-| \sim M$ we can use the $K$-$G$ decomposition 
for the gluon with momentum
$k_1$. The contribution from the $K$ part factorizes and the integral 
over the component $k_1^-$ has the form
\bea \label{factorize}
A_1 & =  & \int \frac{{\mathrm d} k_1^-}{v_A \cdot k_1} \, 
\left[R_A^{ \, a_1 \ldots \, a_n}(k_1^-,k_2^-=0,
\ldots,k_{n-1}^-=0) \, - \, \theta (M - |k_1^-|) \, R_A^{ \, a_1 
\ldots \, a_n}(k_1^-=0, \ldots,k_{n-1}^-=0)\right] \nonumber \\
& \times & \sum_{i=2}^{n-1} \left( i g_s f^{a_1 c_i a_i} \, 
\int_{-M}^{M} \prod_{j=2}^{n-1}
{\mathrm d}k^-_j \, J_A^{(n-1) \, a_2 \ldots \, c_i \ldots \, 
a_n}(p_A, q, \eta, v_B; k_2, \ldots, k_1 + k_i, \ldots, k_n) \right).
\eea
Eq. (\ref{factorize}) is valid when all the lines inside the jet are 
jet-like. In that case the contributions from the ghosts are power 
suppressed.
The contribution corresponding to a $G$ gluon comes from the region 
of integration shown in
Fig. \ref{decoupling}b. It can be expressed in the form of Eq. 
(\ref{minusint}) involving some $J_A^{(n')}$
with fewer loops than in the original $J_A^{(n)}$, and an $R_A'$ with 
more loops than in the original $R_A$.
Then we can repeat the steps described above with this new integral.

Every subsequent term in the square bracket of Eq. (\ref{iden}) can 
be treated the same way as the first term. This allows us to express 
the integral in Eq. (\ref{minusint}) in terms of $k_i^-$ integrals 
over some $J_A^{(n')}$s, which have the same or fewer number of loops 
than the original $J_A^{(n)}$,
\bea \label{minusint2}
& & {\Gamma}_A^{(n') \, a'_1 \ldots \, a'_{n'}} \left( p_A, q, \eta, 
v_B; \, k_1^{' \; +}, \ldots, k_{n'}^{' \; +}; \, k'_{1 \perp},
\ldots, k'_{n' \perp}; M \right) \equiv  \nonumber \\
& & \int_{-M}^{M} \prod_{i=1}^{n'-1} {\mathrm d}k^{' \; -}_i \, 
J_A^{(n') \, a'_1 \ldots \, a'_{n'}}
\left( p_A, q, \eta, v_B; k'_1, \ldots, k'_{n'} \right).
\eea
We now want to set $k_i^{' \; +} =0$ in order to put Eq. 
(\ref{minusint}) into the
form of Eq. (\ref{fact2}). To that end,
we employ an identity for $J_A^{(n')}$ (we again suppress the dependence on the
color indices for brevity)
\bea \label{iden2}
J_A^{(n')}(p_A, q, \eta, v_B; k'_1, \ldots, k'_{n'}) & = &
J_A^{(n')} \left(p_A, q, \eta, v_B; k_1^{' \; +}=0, \ldots, k_{n'}^{' 
\; +} = 0, k_1^{' \; -}, \ldots, k_{n'}^{' \; -},
k'_{1 \perp}, \ldots, k'_{n' \perp} \right) \nonumber \\
& + & \sum_{i=1}^{n'-1} \int_0^{k_i^{' \; +}} {\mathrm d}
l^+_i \frac{\partial}{\partial l^+_i} \, J_A^{(n')}\left( p_A, q, 
\eta, v_B; k'_{1 \perp}, \ldots, k'_{n' \perp}, k_1^{' \; -},
\ldots, k_{n'}^{' \; -}, \right. \nonumber \\
& & \left. k_1^{' \; +}, \ldots, k^{' \; +}_{i-1}, \, l_i^+, 
k_{i+1}^{' \; +}=0, \ldots, k_{n'}^{' \; +}=0 \right).
\eea
Substituting the first term of Eq. (\ref{iden2}) into Eq. 
(\ref{minusint2}), we recognize the definition for ${\Gamma}_A$, Eq. 
(\ref{gammaDef}). We have shown in Sec. \ref{jet3} that the 
contributions from the terms proportional to
$\partial J_A^{(n')} / \partial l^+_i$ in Eq. (\ref{iden2}) can be 
expressed as soft-loop integrals of some $J_A^{(n'')}$, again with 
fewer loops than in $J_A^{(n')}$. When we substitute this into Eq. 
(\ref{minusint2}) we may express the resulting contribution in terms 
of integrals which have the form of Eq. (\ref{minusint}). We can now 
repeat all the steps mentioned so far, with this new integral. By 
this iterative procedure we can transfer the $k_i^-$ integrals
in Eq. (\ref{minusint}) to $J_A^{(n)}$ and also set $k_i^+ = 0$ 
inside $J_A^{(n)}$. In a similar manner, we can analyze
the $p_j^+$ integrals in Eq. (\ref{fact1}), and express them in terms 
of ${\Gamma}_B$ defined in Eq. (\ref{gammaDef}).
This algorithm, indeed, leads from the first factorized form of the 
considered amplitude, Eq. (\ref{fact1}), to the second factorized 
form, Eq. (\ref{fact2}).
   
\subsection{Evolution equation} \label{evoleq}

We have now collected all the ingredients necessary to derive the 
evolution equations for quantities defined in Eq. (\ref{gammaDef}). 
Consider ${\Gamma}_A^{(n)}$. We aim to find an expression for 
$p_A^+ \partial {\Gamma}^{(n)}_A / \partial p_A^+$. As discussed in 
Sec. \ref{jet2} this will enable us to resum the large logarithms of 
$\ln(p_A^+)$ and eventually the logarithms of $\ln(s/|t|)$. According 
to Eq. (\ref{gammaDef}), in order to find $p_A^+ \partial 
{\Gamma}^{(n)}_A / \partial p_A^+$, we need to study $p_A^+ \partial 
J^{(n)}_A / \partial p_A^+$. Using the identities
$p_A \cdot v_B = 2 \, {\xi}_A {\zeta}_B$, $p_A \cdot k_i = 2 \, 
{\xi}_A {\xi}_i$, where ${\xi}_i \equiv k_i^- {\eta}^+$ and
${\xi}_A, {\zeta}_B$ are defined in Eq. (\ref{xidef}), we conclude that
\be
J_A^{(n)} = {\zeta}_B^n \, {\bar J}_A^{(n)}\left({\xi}_A, \, 
\{{\xi}_i\}_{i=1}^{n-1}, \, t, \,
\{q_{\perp} \cdot k_{i\perp}\}_{i=1}^{n-1}, \, \{k_{i\perp} \cdot 
k_{j\perp}\}_{i,j=1}^{n-1}\right).
\ee
 From this structure, using the chain rule, we derive the following 
relation satisfied by $J_A^{(n)}$, which generalizes Eq. 
(\ref{evolut}) to $J^{(n)}_A$ with arbitrary number of external 
gluons,
\be \label{evolJn}
p_A^+ \frac{\partial J_A^{(n)}}{\partial \, p_A^+} = - {\tilde 
\eta}^{\alpha}\frac{\partial J_A^{(n)}}{\partial \, {\eta}^{\alpha}} +
\sum_{i=1}^{n-1}k_i^- \frac{\partial J_A^{(n)}}{\partial k_i^-}
+ {\zeta}_B \frac{\partial J_A^{(n)}}{\partial {\zeta}_B}.
\ee
Now, we integrate both sides of Eq. (\ref{evolJn}) over
$\prod_{j=1}^{n-1} \left( \int_{-M}^{M} \mathrm{d} k^-_j \right )$
and set all $k_j^+=0$. Then, using the definition for
${\Gamma}_A^{(n)}$, Eq. (\ref{gammaDef}), the left hand side is 
nothing else but $p_A^+ \partial
{\Gamma}^{(n)}_A / \partial p_A^+$. The first term on the right hand side of 
Eq. (\ref{evolJn}) is simply $ - {\tilde
\eta}^{\alpha} \partial \, {\Gamma}_A^{(n)} / \partial \, 
{\eta}^{\alpha}$. Noting that ${\zeta}_B
\partial J_A^{(n)}/\partial {\zeta}_B = n \, J_A^{(n)}$, the last 
term gives simply $n \,
{\Gamma}_A^{(n)}$. For the middle term, we use integration by parts
\bea \label{byParts}
& &\prod_{j=1}^{n-1} \left( \int_{-M}^{M} \mathrm{d} k^-_j \right )
\sum_{i=1}^{n-1} \, k_i^- \frac{\partial J_A^{(n)}}{\partial k_i^-} \, = \,
\prod_{j=1}^{n-1} \left( \int_{-M}^{M} \mathrm{d} k^-_j \right )
\sum_{i=1}^{n-1} \left[ \frac{\partial}{\partial k_i^-} ( k_i^- \, 
J_A^{(n)} ) -
J_A^{(n)} \right] = \nonumber \\
& & \sum_{i=1}^{n-1} \int_{-M}^{M} \left(\prod_{j \neq i}^{n-1} { \mathrm d}
k_j^- \right) \, M \, \left[ \, J_A^{(n)}(k_i^- = +M, \ldots) + 
J_A^{(n)}(k_i^- = -M, \ldots) \, \right]
  \; - \; (n-1) \, {\Gamma}_A^{(n)}.
\eea
Combining the partial results, Eqs. (\ref{evolJn}) and 
(\ref{byParts}), we obtain the
following evolution equation
\bea \label{evolG}
p_A^+ \, \frac{\partial \, {\Gamma}_A^{(n)}}{\partial \, p_A^+} & = &
\sum_{i=1}^{n-1} \int_{-M}^{M} \left(\prod_{j \neq i}^{n-1} { \mathrm d}
k_j^- \right) \, M \, \left[ J_A^{(n)}(k_i^- = +M, \ldots) + 
J_A^{(n)}(k_i^- = -M, \ldots) \right]
  \nonumber \\
& & + \, {\Gamma}_A^{(n)} \, - \, {\tilde \eta}^{\alpha} \frac{\partial \,
{\Gamma}_A^{(n)}}{\partial \, {\eta}^{\alpha}}.
\eea
The jet function $J_A^{(n)}$ in the first term of Eq. (\ref{evolG})
is evaluated at $\{k_i^+=0\}_{i=1}^{n}$ and the $k_j^-$s are 
integrated over for $j=1,\ldots,n-1$
and $j \neq i$.
The first term in Eq. (\ref{evolG}) can be analyzed using the $K$-$G$ 
decomposition for gluon $i$
since the $k_i^-$ is evaluated at the scale $M \sim \sqrt{|t|}$. The 
outcome of the last term in Eq.
(\ref{evolG}) has been determined in Sec. \ref{jet2}, Fig. 
\ref{variationJ} \footnote{ Strictly
speaking we have analyzed
${\tilde \eta}^{\alpha} \partial \, J_A^{(n)} / \partial \, 
{\eta}^{\alpha}$, but because of the relationship between $J_A^{(n)}$ 
and ${\Gamma}_A^{(n)}$ given by Eq. (\ref{gammaDef}), once we know 
${\tilde \eta}^{\alpha} \partial \, J_A^{(n)} / \partial \, 
{\eta}^{\alpha}$ we also know
${\tilde \eta}^{\alpha} \partial \, {\Gamma}_A^{(n)} / \partial \, 
{\eta}^{\alpha}$.}.
As a result we have all the tools necessary to determine the 
asymptotic behavior of the high energy amplitude for process 
(\ref{qqqq}).
To demonstrate this, we will rewrite Eq. (\ref{evolG}) into the form 
where on the right hand side there will be a sum of terms involving
${\Gamma}_A^{(n')}$s convoluted with functions which do not depend on 
$p_A^+$. Let us proceed term by term.

Again, the $K$-$G$ decomposition applies to the first term in Eq. (\ref{evolG})
because the external momenta are fixed with
$k_i^- = \pm M$. Using the factorization of a $K$ gluon given in
Eq. (\ref{gy2}) it is clear that the contributions from the $K$ gluons cancel
for $J_A^{(n)}$s evaluated at $k_i^- = + M$ and $k_i^- = - M$. Hence 
only the $G$ gluon contribution
survives in this term. Its most general form is shown in
Fig. \ref{decoupling}b. Before writing it down let us
introduce the following notation. For a set of indices $\{1, 2, 
\ldots , n\} \backslash \{i\}$
consider all the possible subsets of this set, with $1, 2, \ldots, (n 
- 1)$ number of elements. Let
us denote a given subset by $\pi$, its complementary
subset $\bar{\pi}$, the number of elements in this
subset as $n_{\pi}$ and in its complementary as $n_{\bar{\pi}} \equiv 
(n - 1) - n_{\pi}$.  With this
notation, we can write the $i$th contribution to the first term in 
Eq. (\ref{evolG}) in the form
\bea \label{Gcontrib}
  & & J_A^{(n) \, a_1 \, \ldots \, a_n} \left( k_i^- = +M, \ldots 
\right) + J_A^{(n) \, a_1 \, \ldots \, a_n}
\left( k_i^- = -M, \ldots \right) =
\sum_{\pi} \int \prod_{j = 1}^{N - 1} \frac{{\mathrm d}^D l_j}{(2 
\pi)^D} \nonumber \\
  & & S^{\mu_1 \ldots \, \mu_N}_{a_i \, a_{i_1} \ldots \, 
a_{i_{n_\pi}} \, b_1 \ldots \, b_N}
\left( k_i^- = + M, k_{i_1}^-, \ldots, k_{i_{n_{\pi}}}^-; k_i^+ = 0, 
k_{i_1}^+ = 0, \ldots,
k_{i_{n_{\pi}}}^+ = 0; k_{i \perp}, k_{i_1 \, \perp}, \ldots, 
k_{i_{n_{\pi}} \, \perp}; \right. \nonumber \\
& & \left. l_1, \ldots, l_N; q, \eta \right) \, \times \, 
J_{A \; \mu_1 \ldots \, \mu_N}^{(n_{\bar{\pi}} + N) \,
a_{\bar{i}_1} \, \ldots \, a_{\bar{i}_{n_{\bar{\pi}}}} \, b_1 \ldots 
\, b_N} \left (k_{\bar{i}_1}^-,
\ldots, k_{\bar{i}_{n_{\bar{\pi}}}}^-; k_{\bar{i}_1}^+ = 0, \ldots, 
k_{\bar{i}_{n_{\bar{\pi}}}}^+ = 0; k_{\bar{i}_1 \, \perp}, \ldots, 
k_{\bar{i}_{n_{\bar{\pi}}} \, \perp}; \right. \nonumber \\
& & \left. l_1, \ldots, l_N; p_A, q, \eta \right) + (k_i^- 
\rightarrow - M).
\eea
In Eq. (\ref{Gcontrib}), the summation over repeated indices is understood.
We sum over all possible subsets $\pi$. In other words, we sum over 
all possible attachments of
external gluons to jet function $J_A$ and to the soft function $S$. 
The elements of  a given set
$\pi$ are denoted $i_1, i_2, \ldots, i_{n_{\pi}}$. The elements of a 
complementary set $\bar{\pi}$
are labeled
$\bar{i}_1, \bar{i}_2, \ldots, \bar{i}_{n_{\bar{\pi}}}$. The number 
of gluons connecting $S$ and
$J_A^{(n_{\bar{\pi}} + N)}$ is $N$.

Following the procedure described in Sec. \ref{secondFactorizedForm} 
with $R_A$ in Eq. (\ref{minusint}) replaced by $S$ in Eq. 
(\ref{Gcontrib}), we can express the contribution from a $G$ gluon in 
the first term of Eq. (\ref{evolG}) in a form
\bea \label{evol1}
\lefteqn{ \sum_{i=1}^{n-1} \int_{-M}^{M} \left(\prod_{j \neq i}^{n-1} 
{ \mathrm d}
k_j^- \right) \, M \, \left[ J_A^{(n) \, a_1 \, \ldots \, a_n}(k_i^- 
= +M, \ldots) + J_A^{(n) \, a_1 \, \ldots \, a_n}
(k_i^- = -M, \ldots)\right] = } \nonumber \\
& & \sum_{m} \int \prod_{j = 1}^{m} \, {\mathrm d}^{D-2} l_{j \perp} 
\, {\cal K}^{(n, m)}_{a_1 \ldots \, a_n; \, b_1 \ldots \, b_m}(k_{1 
\perp}, \ldots, k_{n \perp}, l_{1 \perp}, \ldots, l_{m \perp}; q, 
\eta; M) \nonumber \\
& \times & \Gamma^{(m) \, b_1 \ldots \, b_m}_A (p_A, q, \eta; l_{1 
\perp}, \ldots, l_{m \perp}; M).
\eea
The function ${\cal K}^{(n, m)}$ does not contain any dependence on $p_A$.
It can contain delta functions setting some of the color indices $b_i$,
as well as transverse momenta $l_{i \perp}$ of $\Gamma_A^{(m)}$ equal 
to color indices $a_i$ and transverse momenta $k_{i \perp}$
of $\Gamma_A^{(n)}$.

Next we turn our attention to the last term appearing in Eq. 
(\ref{evolG}). The contribution to this term has been depicted 
graphically
in Fig. \ref{variationJ}. Consider the term in Fig. 
\ref{variationJ}a. It can be written in a form
\bea \label{virtualVar}
& & \int_{-M}^M \left(\prod_{j = 1}^{n - 1} {\mathrm d} \, k_j^- 
\right) \, \left( \mathrm{Fig}. \,
\ref{variationJ} \mathrm{a} \right) =
\int_{-M}^M \left(\prod_{j = 1}^{n - 1} { \mathrm d} \, k_j^- \right) 
\sum_{{\pi}}
S'_{a_i \, a_{i_1} \ldots \, a_{i_{n_\pi}} \, b} \, (k_i^-, 
k_{i_1}^-, \ldots, k_{i_{n_{\pi}}}^-; \nonumber \\
& & k_i^+ = 0, k_{i_1}^+ = 0, \ldots, k_{i_{n_{\pi}}}^+ = 0; k_{i \, 
\perp}, k_{i_1 \, \perp}, \ldots, k_{i_{n_{\pi}} \, \perp}; l = k_i + 
k_{i_1} + \ldots + k_{i_{n_{\pi}}}; q, \eta) \nonumber \\
& & \times \; J_{A}^{(n_{\bar{\pi}} + 1) \, a_{\bar{i}_1} \, \ldots 
a_{\bar{i}_{n_{\bar{\pi}}}} \, b} (k_{\bar{i}_1}^-, \ldots, 
k_{\bar{i}_{n_{\bar{\pi}}}}^-; k_{\bar{i}_1}^+ = 0, \ldots, 
k_{\bar{i}_{n_{\bar{\pi}}}}^+ = 0; k_{\bar{i}_1 \, \perp}, \ldots, 
k_{\bar{i}_{n_{\bar{\pi}}} \, \perp}; \, \nonumber \\
& & \hspace*{3.5cm} l = k_i + k_{i_1} + \ldots + k_{i_{n_{\pi}}}; 
p_A, q, \eta).
\eea
In Eq. (\ref{virtualVar}), we have used the same notation as in Eq. 
(\ref{Gcontrib}). Momentum $l$ connects $S'$ with
$J_A^{(n_{\bar{\pi}} + 1)}$. Following the same procedure as in Sec. 
\ref{secondFactorizedForm} with $R_A$ appearing in Eq. 
(\ref{minusint}) replaced by $S'$ introduced in Eq. 
(\ref{virtualVar}), we can express this contribution in a form given 
by Eq. (\ref{evol1}) with a different kernel ${\cal K}^{(n, m)}$.

The contribution from Fig. \ref{variationJ}b can be written
\bea \label{realVar}
& & \int_{-M}^{M} \left(\prod_{j = 1}^{n - 1} {\mathrm d} \, k_j^- 
\right) \, \left( \mathrm{Fig}. \,
\ref{variationJ}\mathrm{b} \right)
\, = \, \int_{-M}^{M} \left(\prod_{j = 1}^{n - 1} { \mathrm d} \, 
k_j^- \right) \sum_{{\pi}}
\int \frac{\mathrm{d}^D k}{(2\pi)^D} \, S''_{a_i \, a_{i_1} \ldots \, 
a_{i_{n_\pi}} \, b \, c} \,
(k_i^-, \nonumber \\
& & k_{i_1}^-, \ldots, k_{i_{n_{\pi}}}^-; k_i^+ = 0, k_{i_1}^+ = 0, 
\ldots, k_{i_{n_{\pi}}}^+ = 0; k_{i \, \perp}, k_{i_1 \, \perp},
\ldots, k_{i_{n_{\pi}} \, \perp}; k, l; q, \eta) \, \times \nonumber \\
& & J_{A}^{(n_{\bar{\pi}} + 2) \, a_{\bar{i}_1} \, \ldots \, 
a_{\bar{i}_{n_{\bar{\pi}}}} \, b \, c}
\left(k_{\bar{i}_1}^-, \ldots, k_{\bar{i}_{n_{\bar{\pi}}}}^-; 
k_{\bar{i}_1}^+ = 0, \ldots,
k_{\bar{i}_{n_{\bar{\pi}}}}^+ = 0; k_{\bar{i}_1 \, \perp}, \ldots, 
k_{\bar{i}_{n_{\bar{\pi}}} \, \perp}; k, l; p_A, q, \eta \right).
\eea
The flow of momenta $k$ and $l$ is exhibited in Fig. 
\ref{variationJ}b. The momentum $k$ flows through the boxed vertex and
the ghost line shown in Fig. \ref{variationJ}b which forces this 
momentum to be soft,
so that lines $k$ and $l$
are part of the function $S''$. Since the line with momentum $k$ is 
soft, then all
gluons attaching to $J_{A}^{(n_{\bar{\pi}} + 2)}$ in Eq. 
(\ref{realVar}) are soft and we can again
apply the procedure described in Sec. \ref{secondFactorizedForm} to 
bring the contribution in Fig.
\ref{variationJ}b into the form given by Eq. (\ref{evol1}) with a 
different kernel,  of course.

In summary, we have demonstrated that all the terms on the right hand 
side of Eq. (\ref{evolG}) can
be put into the form given by Eq. (\ref{evol1}). This indicates that 
Eq. (\ref{evolG}), indeed,
describes the evolution of $\Gamma_A^{(n)}$ in $\ln p_A^+$ since it 
can be written as
\bea \label{evol}
\lefteqn{ \left(p_A^+ \, \frac{\partial}{\partial \, p_A^+} - 1 
\right) \, {\Gamma}_A^{(n) \; a_1 \ldots \, a_n}(p_A, q, \eta;
k_{1 \perp}, \ldots, k_{n \perp}) = } \nonumber \\
& & \sum_{m} \int \prod_{j = 1}^{m} \, {\mathrm d}^{D-2} l_{j \perp} 
\, {\cal K}^{(n, m)}_{a_1 \ldots \, a_n; \, b_1 \ldots \, b_m}(k_{1 
\perp}, \ldots, k_{n \perp}, l_{1 \perp}, \ldots, l_{m \perp}; q, 
\eta) \nonumber \\
& & \times \, \Gamma^{(m) \; b_1 \ldots \, b_m}_A(p_A, q, \eta; l_{1 
\perp}, \ldots, l_{m \perp}).
\eea
The kernels ${\cal K}^{(n, m)}$ do not depend on $p_A^+$.  As 
indicated above, they can contain delta
functions setting  some of the color indices $b_i$, as well as 
transverse momenta $l_{i \perp}$ of
$\Gamma_A^{(m)}$ equal to color indices
$a_i$ and transverse momenta $k_{i \perp}$ of $\Gamma_A^{(n)}$. The 
systematic use of this evolution equation enables us to resum
large logarithms $\ln (p_A^+)$ at arbitrary level of logarithmic 
accuracy. Analogous equation is satisfied by $\Gamma_B$. It resums 
logarithms of $\ln(p_B^-)$.

\subsection{Counting the number of logarithms} \label{doubleLogCancellation}

Having derived the evolution equations for ${\Gamma}_A^{(n)}$, Eqs. 
(\ref{evolG}) and (\ref{evol}), it does not take too much effort to 
show that at $r$-loop order the amplitude contains at most $r$ powers 
of $\ln(s/|t|)$. We follow the method of Ref. \cite{sen83}. We have 
argued in Sec. \ref{secondFactorizedForm} that the power of 
$\ln(s/|t|)$ in the overall amplitude corresponds to the power of 
$\ln(p_A^+)$ in ${\Gamma}_A^{(n)}$. So we have to demonstrate that at 
$r$-loop order ${\Gamma}_A^{(n,r)}$, where ${\Gamma}_A^{(n,r)}$ 
represents a contribution to ${\Gamma}_A^{(n)}$ at $r$-loop level, 
does not contain more than $r$ logarithms of $\ln(p_A^+)$. We prove 
this statement by induction. First of all, the tree level 
contribution to ${\Gamma}_A^{(n,0)}$ is proportional to the expression
\be \label{treeG}
\int_{-M}^M \left( \prod_{i=1}^{n-1} {\mathrm d} k_i^- \right)
\sum_{\{i_1, \ldots, i_n\}} \, \prod_{j=1}^{n-1} \, \frac{1}{(p_A - 
\sum_{l=1}^{j} k_{i_l})^2 + i\epsilon} \,
\left( \prod_{j=n}^{1} \, t^{a_{i_j}} \right)_{r_1, r_A},
\ee
where $t^{a_{i_j}}$s are the generators of the $SU(3)$ algebra in the 
fundamental representation. The sum over $\{i_1,\ldots,i_n\}$ 
indicates that we sum over all possible insertions of the external 
soft gluons. Eq. (\ref{treeG}) is evaluated at $\{k_i^+=0\}_{i=1}^n$. 
Expanding the denominators in Eq. (\ref{treeG}) we obtain the 
expression $-2 p_A^+ (k_{i_1}^- + \ldots + k_{i_j}^-) - (k_{i_1} + 
\ldots + k_{i_j})^2_{\perp} + i\epsilon$. We see that the poles in 
$k_i^-$ planes are not pinched and therefore the $k_i^-$ integrals 
cannot produce $\ln(p_A^+)$ enhancements.

Next we assume that the statement is true at $r$-loop order, and show that
it then also holds at $(r+1)$-loop level. To this end we consider the 
evolution equation, Eq.
(\ref{evolG}), and  examine $(p_A^+ \, \partial/\partial \, p_A^+ -1) 
\, {\Gamma}_A^{(n,r+1)}$. Its
contribution is given by the first and the third term on the right 
hand side of Eq. (\ref{evolG}).
As already mentioned, the first term in  Eq. (\ref{evolG}) can be 
analyzed using $K$-$G$
decomposition. The contributions from the $K$ terms cancel each other 
while the contribution from the
$G$ gluons are given by the kind of diagram shown in Fig. 
\ref{decoupling}b. The latter, however,
can be written as a sum of soft loop integrals over
$J_A^{(n',r')}$ with $r' \le r$, since we loose at least one loop in 
the original $J_A^{(n, r+1)}$
due to the soft momentum integration. This is demonstrated in  Eq. 
(\ref{Gcontrib}). Following the
procedure described in Sec. \ref{secondFactorizedForm}, we may 
express these contributions as
transverse momentum integrals of some ${\Gamma}_A^{(n',r')}$, see Eq. 
(\ref{evol1}). These contain
at most $r'\le r$ logarithms of $\ln(p_A^+)$. The contribution from 
the third term in the evolution
equation, Eq. (\ref{evolG}), is given by the diagrams depicted in 
Fig. \ref{variationJ}. These are
again soft loop integrals of some $J_A^{(n',r')}$ with $r' \le r$, 
and they can be expressed as
transverse momentum integrals of ${\Gamma}_A^{(n',r')}$, see Eqs. 
(\ref{virtualVar}) and
(\ref{realVar}), which have, therefore, at most $r$ logarithms of 
$\ln(p_A^+)$. Since both terms on
the right hand side of Eq. (\ref{evolG}) have at most $r$ logarithms 
of $\ln(p_A^+)$, then also
$p_A^+ \partial \, {\Gamma}_A^{(n,r+1)} / \partial \, p_A^+$ has at 
most $r$ logarithms of
$\ln(p_A^+)$ at $(r+1)$-loop level. This immediately shows that 
${\Gamma}_A^{(n,r+1)}$ itself cannot
have more than $(r+1)$ logarithms of $\ln(p_A^+)$ at $(r+1)$-loop 
level.  

This result enables us to formally classify the types of diagrams 
which contribute to the amplitude at the $k$-th nonleading 
logarithm level. As has been shown in Sec. 
\ref{secondFactorizedForm}, we can write an arbitrary
contribution to the amplitude for process (\ref{qqqq}) in the Regge 
limit in the second factorized
form given by Eq. (\ref{fact2}). Consider an $r$-loop contribution 
to the amplitude and let $L_A$,
$L_B$ and $L_S$ be the number of loops contained in $\Gamma_A$, 
$\Gamma_B$ and $S$. Since $\Gamma_A$
($\Gamma_B$) can contain $L_A$ ($L_B$) number of logarithms of 
$p_A^+$ ($p_B^-$) at most, the
maximum number of logarithms, $N_{\mathrm{maxLog}}$, we can get is
\be \label{logCount}
N_{\mathrm{maxLog}} = r - L_S.
\ee
This indicates that when evaluating the amplitude at the $k$-th 
nonleading approximation, we need to consider diagrams where $1, 2, 
\ldots, (k+1)$ soft gluons are exchanged between the jet functions 
$J_A$ and $J_B$.

\subsection{Solution of the evolution equations} \label{solEvolEq}

Having obtained the evolution equations, Eqs. (\ref{evolG}) and
(\ref{evol}), we discuss how to construct their solution. 
Our starting point is Eq. (\ref{evol}). In shorthand notation it reads
\bea \label{evolr}
p_A^+ \, \frac{\partial}{\partial \, p_A^+} \, {\Gamma}_A^{(n, r)} =
\sum_{r'=0}^{r - 1} \sum_{n'} \, 
{\cal K}^{(n, n'; r - r')} \otimes \, \Gamma_A^{(n', r')},
\eea
at $r$-loop level. Indices $n$ and $n'$, besides denoting the number of
external gluons of the jet function, 
also label the transverse momenta and the color
indices of these gluons. The symbol $\otimes$ in Eq. (\ref{evolr}) denotes
convolution over the transverse momenta and the color indices.
Note that Eq. (\ref{evolr}) holds for $\Gamma_A$ with the overall factor
$p_A^+$ divided out $(\Gamma_A \equiv \Gamma_A / p_A^+)$. 
We have proved, in Sec. \ref{doubleLogCancellation}, that
$\Gamma_A^{(n,r)}$ can contain at most $r$ logarithms of $\ln(p_A^+)$ at 
$r$-loop level. Therefore the most general expansion for $\Gamma_A$ is
\be \label{gammaExpand}
\Gamma_A^{(n,r)} \equiv \sum_{j = 0}^{r} c_j^{(n, r)} \, \ln^j(p_A^+).
\ee
If we want to know $\Gamma_A^{(n,r)}$ at ${\rm N}^k$LL accuracy ($k = 0$ is
LL, $k = 1$ is NLL, etc.), we need to find all
$c_j^{(n,r)}$ such that $r - j \le k$.     
The coefficients $c_j^{(n,r)}$ in Eq. (\ref{gammaExpand}) depend on the
transverse momenta and the color indices of the 
external gluons. Using the expansion for $\Gamma_A^{(n,r)}$ and
$\Gamma_A^{(n',r')}$, Eq. (\ref{gammaExpand}), in Eq. (\ref{evolr}) 
and comparing the coefficients with the same power of $\ln(p_A^+)$, we
obtain the recursive relation satisfied by the 
coefficients $c_j^{(n,r)}$
\be \label{recurC}
j \, c_j^{(n,r)} = \sum_{r' = j-1}^{r-1} \sum_{n' = 1}^{n+r-r'} {\cal
K}^{(n,n';r-r')} \otimes c_{j-1}^{(n',r')}.
\ee
In Eq. (\ref{recurC}), we have used that, in general, $1 \le n' \le n + r
- r'$.
 
We now show that Eq. (\ref{recurC}) enables us to determine all the
relevant coefficients $c_j^{(r,n)}$ of $\Gamma_A^{(n)}$
order by order in perturbation theory at arbitrary logarithmic accuracy.
We start at LL, $k = 0$, and consider $n = 1$. At $r$-loop level we need
to find the coefficient $c_r^{(1,r)}$. It can 
be expressed in terms of lower loop coefficients using Eq. (\ref{recurC})
and setting $j = r$ and $n = 1$
\be \label{coeff1}
r \, c_r^{(1,r)} = \sum_{n' = 1}^{2} {\cal K}^{(1,n';1)} \otimes
c_{r-1}^{(n',r-1)}.
\ee
In Sec. \ref{llAmplitude} we will prove that the one loop kernel satisfies
${\cal K}^{(1,2;1)} = 0$, Eq. (\ref{k1}). 
This implies that in Eq. (\ref{coeff1}) the coefficient $c_r^{(1,r)}$ is
expressed in terms of lower loop 
coefficient $c_{r-1}^{(1,r-1)}$ and hence, we can construct the
coefficients at arbitrary loop level once 
we compute $c_0^{(1,0)}$, the coefficient corresponding to the tree level
jet function $\Gamma_A^{(1,0)}$.   

Next we construct all $\Gamma_A^{(n)}$ for $n > 1$ at LL accuracy. Let us
assume that we know all $c_{r}^{(n',r)}$ for all $r$ 
and for $n' < n$. We apply Eq. (\ref{recurC}) for $j = r$
\be \label{coeffr}
r \, c_r^{(n,r)} = \sum_{n' = 1}^{n+1} {\cal K}^{(n,n';1)} \otimes
c_{r-1}^{(n',r-1)}.
\ee
In Sec. \ref{nllAmplitude} we will show that the evolution kernel in Eq.
(\ref{coeffr}) obeys 
${\cal K}^{(n,n';1)} = \theta(n-n') \, {\tilde {\cal K}}^{(n,n';1)}$,
Eq. (\ref{ktheta}), where $\theta(n-n')$ is the step function.
This implies that the sum over $n'$ in Eq. (\ref{coeffr}) terminates at
$n' = n$. 
Isolating this term in Eq. (\ref{coeffr}), we can write
\be \label{coeff2}
r \, c_r^{(n,r)} = {\cal K}^{(n,n;1)} \otimes c_{r-1}^{(n,r-1)} + \sum_{n'
= 1}^{n-1} {\cal K}^{(n,n';1)} \otimes c_{r-1}^{(n',r-1)}.
\ee
So after we calculate the tree level coefficient $c_0^{(n,0)}$, we can
construct all the coefficients 
$c_r^{(n,r)}$ using Eq. (\ref{coeff2}) order by order in perturbation
theory, since according to the
assumption we know
$c_{r}^{(n',r)}$ for all $r$ and for $n' < n$. This proves that we can
construct the jet functions at LL, $k = 0$, 
for all $n$ to all loops.  

We now assume that we have constructed all the jet functions at the ${\rm
N}^k$LL accuracy for a
given
$k \ge 0$ and we will show that  we can determine all the jet functions at
the ${\rm N}^{k+1}$LL
level. We start with $n = 1$. Using Eq. (\ref{recurC})  with $n = 1$, $j =
r - (k+1)$, isolating the
term with $r' = r-1$ in the sum over $r'$ and using
${\cal K}^{(1,n';1)} = \delta_{1 \, n'} \, {\cal K}^{(1,1;1)}$, we arrive
at
\be \label{coeff3}
(r-k-1) \, c_{r-k-1}^{(1,r)} = {\cal K}^{(1,1;1)} \otimes
c_{r-k-2}^{(1,r-1)} + 
\sum_{r' = r-k-2}^{r-2} \sum_{n' = 1}^{1+r-r'} 
{\cal K}^{(1,n';r-r')} \otimes c_{r-k-2}^{(n',r')}.
\ee
After we evaluate the coefficient $c_{0}^{(1,k+1)}$ (impact factor), Eq.
(\ref{coeff3}) implies that
we can calculate  the coefficients $c_{r-k-1}^{(1,r)}$ order by order in
perturbation theory,
because, according to the induction assumption,  we know all the
coefficients $c_{r-k-2}^{(n',r')}$
since they are at most ${\rm N}^k$LL.  Once the coefficients of
$\Gamma_A^{(1)}$ are determined at 
${\rm N}^{k+1}$LL level, we assume that we know all the coefficients of
$\Gamma_A^{(n')}$s 
for $n' < n$. We want to show
that we can now construct all the coefficients for $\Gamma_A^{(n)}$ at
${\rm N}^{k+1}$LL accuracy.
First we need to calculate
$c_0^{(n,k+1)}$. Then we use Eq. (\ref{recurC}) to express the coefficient
$c_{r-k-1}^{(n,r)}$, isolating the terms with
$r' = r - 1$ and $n' = n$, as
\bea \label{coeff4}
(r-k-1) \, c_{r-k-1}^{(n,r)} & = & {\cal K}^{(n,n;1)} \otimes
c_{r-k-2}^{(n,r-1)} + 
\sum_{n' = 1}^{n-1} {\cal K}^{(n,n';1)} \otimes c_{r-k-2}^{(n',r-1)}
\nonumber \\
& + & \sum_{r' = r-k-2}^{r-2} \sum_{n' = 1}^{n+r-r'} {\cal
K}^{(n,n';r-r')} \otimes c_{r-k-2}^{(n',r')}.
\eea
The terms appearing in the sum over $r'$ in Eq. (\ref{coeff4}) are known
according to the assumptions 
since for them $r' - (r - k - 2) \le k$. We also know, according to the
induction assumptions, 
the contributions to the second term of Eq. (\ref{coeff4}), since they
have $n' < n$. 
Therefore, we can construct $c_{r-k-1}^{(n,r)}$
order by order in perturbation theory. This finishes our proof that we can
determine the high energy behavior of 
$\Gamma_A^{(n)}$ at arbitrary logarithmic accuracy.  Note that to any
fixed accuracy only a finite number
of fixed-order calculations of kernels and  coefficients $c_0^{(n,r)}$
must be carried out.
In a similar way we can construct a solution for $\Gamma_B^{(m)}$.

Once we know the high energy behavior for $\Gamma_A^{(n)}$ and
$\Gamma_B^{(m)}$, 
then the second factorized form, Eq. (\ref{fact2}), 
implies that we also know the high energy behavior for the overall
amplitude. 
Because a jet function $\Gamma^{(n)}$ is always
associated with at least $n-1$ soft loop momentum integrals in the amplitude, we infer from
Eq. (\ref{logCount}) 
that if we want to know
this amplitude at ${\rm N}^{K}$LL accuracy, it is sufficient to know
$\Gamma_A^{(n)}$ ($\Gamma_B^{(m)}$) at 
${\rm N}^{K+1 - n}$LL (${\rm N}^{K+1 - m}$LL) level for $n \le K + 1$ ($m
\le K + 1$). 
We note, however, that to construct these functions according to the
algorithm above,
it may be necessary to go to slightly larger, although always finite,
values of $n$ and $m$.
Let us describe how this comes about, starting with the basic
recursion relations for coefficients,  Eq. (\ref{recurC}).  

We assume that for fixed $n$ on
the left-hand side of Eq. (\ref{recurC}), the logarithmic accuracy $k$ is
bounded by
the value necessary to determine the 
overall amplitude to $K$th nonleading logarithm:
$k=r-j\le K+1-n$, which we may rewrite as $n + r - (K+1) \le j \le r$.
On the right-hand side of Eq.\ (\ref{recurC}) we encounter the coefficients of the jet functions
with $n'$ external lines,
satisfying the inequality 
$n' \le n + r - r' \le n + r - (j - 1)$. 
Combining these two inequalities, we immediately obtain 
that $n' \le K + 2$.  Then, for any given number of external gluons $n'$
on the right-hand side, 
we encounter a level of logarithmic accuracy $k'=r' - (j-1) \le n + r - n'
-(j-1) \le K + 2 - n'$. 
This reasoning indicates that, in general, we will
need all $\Gamma_A^{(n')}$ ($\Gamma_B^{(m')}$) at ${\rm N}^{K + 2 - n'}$LL
(${\rm N}^{K + 2 - m'}$LL) level 
for $n' \le K+2$ ($m' \le K+2$), when evaluating the amplitude at ${\rm
N}^{K}$LL accuracy.  
We note that for fermion exchange in QED it was shown in Ref. \cite{sen83}
that only contributions with $n'\le K+1$ are nonzero, 
but for QCD, two-loop calculations appear to indicate, Ref. \cite{tibor}, 
that QCD requires the full range of $n'$ identified above, starting at NLL.

\section{High energy behavior of the amplitude} \label{amplitude}

In the previous sections we have developed the general formalism for
obtaining the high-energy behavior of the scattering amplitude for 
process (\ref{qqqq}) at arbitrary
logarithmic accuracy. In the following subsections we apply these 
techniques to study this amplitude at
LL and NLL level.

\subsection{Amplitude at LL} \label{llAmplitude}

\begin{figure} \center
\includegraphics*{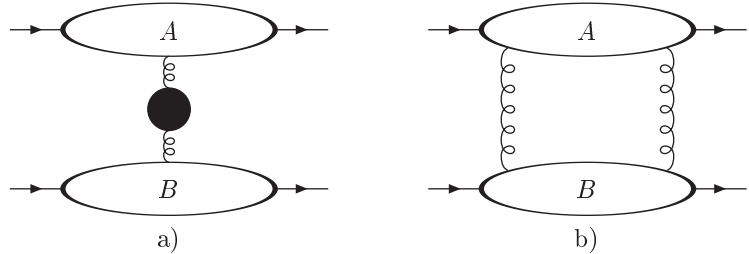}
\caption{\label{nllContribution} Diagrams contributing to the 
amplitude at NLL approximation:
factorized one gluon exchange diagram (a) and nonfactorized two gluon 
exchange diagram (b).}
\end{figure}

\begin{figure} \center
\includegraphics*{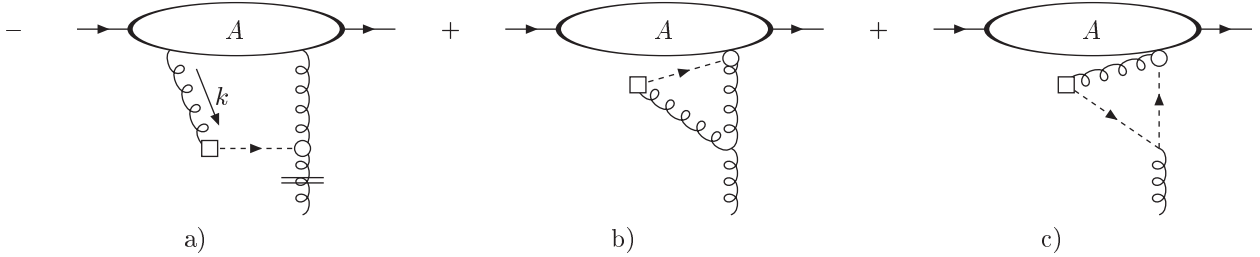}
\caption{\label{lleq} Diagrammatic representation of the evolution 
equation for jet $J_A^{(1)}$ at LL.}
\end{figure}

According to Eq. (\ref{logCount}), the amplitude at LL comes solely 
from the factorized diagram
shown in Fig. \ref{nllContribution}a, but without any gluon 
self-energy corrections. The jet $A$,
containing lines moving in the plus direction, and jet $B$, 
consisting of lines moving in the minus
direction, interact via the exchange of a single soft gluon. This 
gluon couples to jet $A$ via the $-$
component of its polarization and to jet $B$ via the $+$ component of 
its polarization. Since
$v_A^{\alpha} \, N_{\alpha \, \beta}(q, \eta) \, v_B^{\beta} = 1$, 
we can write at LL
\be
A_{\bf 8} \, b_{\bf 8} = - \frac{1}{t} J_{A}^{(1) \, a} (p_A, q, \eta) \,
J_{B}^{(1) \, a} (p_B, q, \eta),
\ee
where $b_{\bf 8}$ is the color basis vector corresponding to the 
octet exchange, defined in Eq. (\ref{qqBasis}).
Using $s = 2p_A^+ \, p_B^-$, the logarithmic derivative of the 
amplitude can be expressed as
\be \label{ev}
\frac{\partial A_{\bf 8}}{\partial \ln s} \, b_{\bf 8} = -\frac{1}{t}
\frac{\partial J_A^{(1) \, a}}{\partial \ln p_A^+} \, J_B^{(1) \, a}
  = -\frac{1}{t} \, J_A^{(1) \, a} \, \frac{\partial J_B^{(1) \, 
a}}{\partial \ln p_B^-} \; .
\ee
In Sec. \ref{jet2}, Eq. (\ref{evolut}), we have derived an evolution 
equation resumming
$\ln(p_A^+)$ in $J_A^{(1)}$.  We note that $J_A^{(1)}=\Gamma_A^{(1)}$, and that
(\ref{evolut}) is a special case of the evolution equation (\ref{evolG}).
The diagrammatic representation of the first term on the far right 
hand side of Eq. (\ref{evolut}), which follows from
Fig. \ref{variationJ} in the case when we have one external soft 
gluon attached to a jet function,
is given by the diagrams in Fig. \ref{lleq}. Diagram in Fig. 
\ref{lleq}a corresponds to Fig. \ref{variationJ}b and the
diagrams in Figs. \ref{lleq}b and \ref{lleq}c correspond to Fig. 
\ref{variationJ}a for $n=1$.
The diagrams in Figs. \ref{lleq}b and \ref{lleq}c are in the 
factorized form, while the one in Fig. \ref{lleq}a is not.

As discussed in Sec. \ref{jet2}, power counting shows that the loop 
momentum $k$
in Fig. \ref{lleq}a must be soft. This implies that we can make the 
following approximations.
First, since at LL all internal lines of the
jet $A$ are collinear to the $+$ direction, we can neglect the $k^+$ 
dependence of $J_A^{(2)}$,
i.e. we may set $k^+ = 0$
inside $J_A^{(2)}$. Also, we can pick the plus components of the 
vertices where the soft
gluons attach to the jet $J_A^{(2)}$.
A short calculation, which uses the Feynman rules for special lines 
and vertices listed in
Appendix \ref{feynmanRules},
gives the contribution to Fig. \ref{lleq}a in a form
\bea \label{eikonalContrib1}
\mathrm{Fig.} \; \ref{lleq} \mathrm{a} & = & - {\bar g}_s \, t \, 
f_{acb} \, \int \frac{{\mathrm d}^D k}{(2 \pi)^D} \,
\frac{1}{k^2 (k-q)^2
\, k \cdot {\bar k}} \, v_A^{\rho} N_{\rho \mu}(k) S^{\mu}(k) \; 
v_B^{\alpha} N_{\alpha \nu} (q-k) v_A ^{\nu}
\nonumber \\
& \times & v_B^{\beta} \, v_B^{\gamma} \; J_{(A) \; \beta \, 
\gamma}^{(2) \, bc}
\left(p_A, q, \eta; k^+=0, k^-, k_{\perp}\right),
\eea
\begin{figure} \center
\includegraphics*{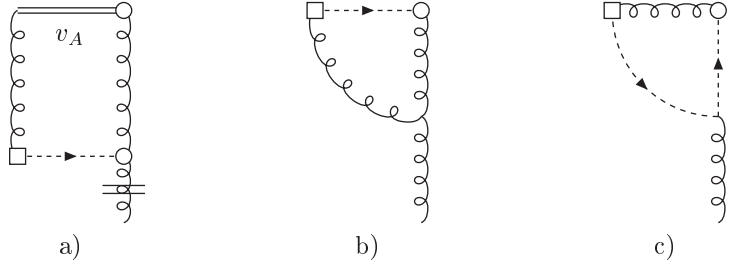}
\caption{\label{trajll} Diagrams determining the contributions to the 
gluon trajectory at the order ${\alpha}_s$.}
\end{figure}
where we have defined ${\bar g}_s \equiv g_s {\mu}^{\epsilon}$.
Using Eqs. (\ref{propagComp}) and (\ref{sCompon}) for the components 
of the gluon
propagator and the boxed vertex,
respectively, it is easy to see that in the Coulomb (Glauber) region, 
$k^- \ll k^+ \sim k_{\perp}$,
the integrand in Eq. (\ref{eikonalContrib1}) becomes an antisymmetric 
function of $k^+$ and
that therefore the integration over $k^+$ vanishes in this region.

In the soft region, where all the components of soft momenta are of 
the same size $\sqrt{-t}$, we can use the $K$-$G$ decomposition for 
the soft gluon with momentum $k$ attached to $J_A^{(2)}$. At LL, 
however, there cannot be any soft internal lines in $J_A^{(2)}$ in 
Eq. (\ref{eikonalContrib1}),
since, as discussed in Sec. \ref{doubleLogCancellation}, only 
integrals over collinear momenta can produce powers of $\ln p_A^+$.
Therefore, at LL, only the $K$ gluon contributes, because the $G$ gluon
must be attached to a soft line.
The $K$ gluon can be decoupled from the rest of the jet $J^{(2)}_A$ 
using the Ward identities, Eq. (\ref{gy2}).
Their application in Eq. (\ref{eikonalContrib1}) gives
\bea \label{eikonalContrib2}
\mathrm{Fig.} \; \ref{lleq} \mathrm{a} & = & - i {\bar g}_s^2 C_A t 
\, \int \frac{{\mathrm d}^D k}{(2 \pi)^D} \,
\frac{1}{k^2 (k-q)^2 \, k \cdot {\bar k} \, v_A \cdot k} \, 
v_A^{\rho} N_{\rho \mu}(k) S^{\mu}(k) \;
v_B^{\alpha} N_{\alpha \nu} (q-k) v_A ^{\nu} \nonumber \\
& \times & J_{A}^{(1) \, a}(p_A, q, \eta).
\eea
We have used the identity $f_{acb} \, f_{dcb} = N_c \, {\delta}_{ad} 
\equiv C_A \, {\delta}_{ad}$ in Eq. (\ref{eikonalContrib2}).
Eq. (\ref{eikonalContrib2}) now gives a factorized form for Fig. \ref{lleq}a.
Since the contributions in Figs. \ref{lleq}b and \ref{lleq}c are 
already in the factorized form,
we can immediately infer that the gluon reggeizes
at LL. Combining the terms from Fig. \ref{lleq} in Eq. 
(\ref{evolut}), we obtain the evolution equation at leading logarithm
\be \label{ev1}
p_A^+ \frac{\partial}{\partial p_A^+} J_A^{(1) \, a}(p_A, q, \eta) = 
\alpha(t) \, J_A^{(1) \, a}(p_A, q, \eta).
\ee
Using the notation for evolution kernels introduced in Sec. \ref{solEvolEq}, 
Eq. (\ref{ev1}) implies that
\be \label{k1}
{\cal K}^{(1,2;1)} = 0.
\ee
In Eq. (\ref{ev1})
\be \label{trajectory}
\alpha (t) \equiv 1 + {\alpha}^{(1)}_{a}(t) + {\alpha}^{(1)}_{b}(t) + 
{\alpha}^{(1)}_{c}(t),
\ee
is the gluon trajectory up to the order ${\alpha}_s$, and 
${\alpha}^{(1)}_a (t)$, ${\alpha}^{(1)}_b (t)$ and ${\alpha}^{(1)}_c 
(t)$ are
its contributions given in Figs. \ref{trajll}a - \ref{trajll}c, respectively,
\bea \label{al1}
{\alpha}^{(1)}_{a}(t) & \equiv & - i {\bar g}_s^2 C_A t \, \int 
\frac{{\mathrm d}^D k}{(2 \pi)^D} \,
\frac{1}{k^2 (k-q)^2 \, k \cdot {\bar k} \, v_A \cdot k} \, 
v_A^{\rho} N_{\rho \mu}(k) S^{\mu}(k) \;
v_B^{\beta} N_{\beta \nu} (q-k) v_A ^{\nu}, \nonumber \\
{\alpha}^{(1)}_{b}(t) & \equiv & i {\bar g}_s^2 C_A \, \int 
\frac{{\mathrm d}^D k}{(2 \pi)^D} \, \frac{1}{k^2 (k-q)^2 \,
k \cdot {\bar k}} \, S_{\alpha}(k) N^{\alpha \, \mu}(k) \; v_A^{\rho} 
N_{\rho}^{\;\; \nu} (q-k) \;
V_{\mu \beta \nu}(k, -q, q-k) v_B^{\beta},\nonumber \\
{\alpha}^{(1)}_{c}(t) & \equiv & - i {\bar g}_s^2 C_A \, \int 
\frac{{\mathrm d}^D k}{(2 \pi)^D} \,
\frac{1}{k^2 \, k \cdot {\bar k} \, (q-k) \cdot ({\bar q}-{\bar k})} 
\, v_A^{\rho} N_{\rho \mu}(k) S^{\mu}(k) \;
(v_B \cdot {\bar k}) \, .
\eea

In Eq. (\ref{al1}), $V_{\mu \beta \nu}(k, -q, q-k)$ stands for the 
momentum part
of the three-point gluon vertex.
After contracting the tensor structures in Eq. (\ref{al1}), using the explicit
form for $V_{\mu \beta \nu}$, $v_A$, $v_B$, $S^{\mu}$
(Eq. (\ref{boxedVertex})) and for the components of the gluon propagator,
Eq. (\ref{propagComp}), we obtain for ${\alpha}^{(1)}_{a,b,c}(t)$,
\bea \label{al2}
{\alpha}^{(1)}_{a} (t) & = & - i {\bar g}_s^2 C_A \frac{t}{2} \, \int 
\frac{{\mathrm d}^D k}{(2 \pi)^D}
\frac{[k_{\perp}^2 k_0 + k^2 k_3][(k-q)^2 - (k-q)^2_{\perp}]}{(k_0 + 
k_3) \, k^2 \, (k-q)^2 \,
(k \cdot {\bar k})^2 \, (k-q) \cdot ({\bar k} - {\bar q})} \; , \nonumber \\
{\alpha}^{(1)}_{b} (t) & = & i {\bar g}_s^2 C_A \frac{1}{2} \, \int 
\frac{{\mathrm d}^D k}{(2 \pi)^D}
\frac{1}{k^2 \, (k-q)^2 \, (k \cdot {\bar k})^2 \, (k-q) \cdot ({\bar 
k} - {\bar q})} \nonumber \\
& \times & [k_{\perp}^2 \, {\bar k}^2 \, (k-q)^2 + 2 k^2 \, k_3^2 \, 
(k-q)_{\perp} \cdot q_{\perp}
+ 2 k_0^2 \, k_3^2 \, k_{\perp} \cdot (k-q)_{\perp} + 2 k_0^2 \, 
k_{\perp}^2 \, (k-q)_{\perp}^2] \; , \nonumber \\
{\alpha}^{(1)}_{c} (t) & = & i {\bar g}_s^2 C_A \frac{1}{2} \, \int 
\frac{{\mathrm d}^D k}{(2 \pi)^D}
\frac{k_3^2}{(k \cdot {\bar k})^2 \, (k-q) \cdot ({\bar k} - {\bar q})} \; .
\eea
Next, we perform the $k^0$ and $k^3$ integrals in Eq. (\ref{al2}). 
For ${\alpha}^{(1)}_a(t)$, these integrals are UV/IR finite.
However in the case of ${\alpha}^{(1)}_{b,c}(t)$, the $k^0$ integral 
is linearly UV divergent.
In order to regularize this energy integral, we invoke split 
dimensional regularization introduced in Ref. \cite{leibbrandt96}.
The idea is to regularize separately the energy and the spatial 
momentum integrals, i.e. to write ${\mathrm d}^4 k_E \rightarrow
{\mathrm d}^{D_1} k_4 \, {\mathrm d}^{D_2} {\vec k}$ for Euclidean 
loop momenta $k_E$. The dimensions $D_1$ and $D_2$ are given by
$D_1 = 1 - 2 {\varepsilon}_1$ and $D_2 = 3 - 2 {\varepsilon}_2$, with 
${\varepsilon}_j \rightarrow 0+$ for $j = 1, 2$.
Since the energy integral for ${\alpha}^{(1)}_c(t)$ is scaleless, it 
vanishes in this split dimensional regularization.
The energy integrals in ${\alpha}^{(1)}_{a,b}(t)$ are straightforward.

All the $k^3$ integrals can be expressed as derivatives
with respect to $k_{\perp}^2$ and/or $(k-q)_{\perp}^2$ of a single integral
\be
I(a,b) \equiv \int_{0}^{\infty} {\mathrm d} k^3 \frac{1}{\sqrt{k^2_3 
+ a^2} \; (k^2_3 + b^2)} = \frac{1}{b\sqrt{b^2 - a^2}} \,
\ln \left(\frac{b + \sqrt{b^2 - a^2}}{a}\right).
\ee
The result of these integrations over $k^3$ is
\bea \label{al3}
{\alpha}^{(1)}_{a} (t) & = & {\alpha}_s {\mu}^{2 \epsilon} \, C_A \, 
t  \, \int \frac{{\mathrm d}^{D-2} k_{\perp}}{(2 \pi)^{D-2}}
\left(I(|k_{\perp}|, |k_{\perp}-q_{\perp}|) \, 
\frac{k_{\perp}^2}{[(k-q)_{\perp}^2 - k_{\perp}^2]^2} +
\frac{2 (k - q)_{\perp}^2 - 3 k_{\perp}^2}{k_{\perp}^2 \, 
[(k-q)_{\perp}^2 - k_{\perp}^2]^2} \right), \nonumber \\
{\alpha}^{(1)}_{b} (t) & = & - {\alpha}_s {\mu}^{2 \epsilon} \, C_A 
\, t  \, \int \frac{{\mathrm d}^{D-2} k_{\perp}}{(2 \pi)^{D-2}}
\left(I(|k_{\perp}|, |k_{\perp}-q_{\perp}|) \, 
\frac{k_{\perp}^2}{[(k-q)_{\perp}^2 - k_{\perp}^2]^2} -
\frac{1}{[(k-q)_{\perp}^2 - k_{\perp}^2]^2} \right), \nonumber \\
{\alpha}^{(1)}_{c} (t) & = & 0.
\eea
Combining the results of Eq. (\ref{al3}) and Eq. (\ref{trajectory}), 
we obtain the standard expression for the gluon trajectory at LL
\be \label{trajec1}
\alpha (t) = 1 + C_A {\alpha}_s {\mu}^{2 \epsilon} \, \int
\frac{{\mathrm d}^{D-2} k_{\perp}}{(2 \pi)^{D-2}} \frac{t}{k_{\perp}^2 \,
(k-q)_{\perp}^2} \; .
\ee
We can now simply solve the evolution equation (\ref{ev}), to derive 
the factorized (reggeized)
form for the amplitude in the color octet
\be
A_{\bf 8}(s,t,{\alpha}_s) = s^{\alpha (t)} \, {\tilde A}_{\bf 8} (t, 
{\alpha}_s).
\ee
The amplitude factorizes into the universal factor $s^{\alpha (t)}$, 
which is common for all processes involving two partons
in the initial and final state and dominated by the gluon exchange, 
and the part ${\tilde A}_{\bf 8}$, the so-called impact factor, which 
is specific to the process under consideration.

\subsection{Amplitude at NLL} \label{nllAmplitude}

At NLL level the contribution to the amplitude comes from both the 
one gluon exchange diagram,
Fig. \ref{nllContribution}a, and from the two gluon exchange diagram, 
Fig. \ref{nllContribution}b.
At this level, both singlet and octet color exchange are possible in 
the latter.
Including the self-energy corrections to the propagator of the 
exchanged gluon (taking into account
the corresponding counter-terms), we can write the contribution from 
the diagram in Fig. \ref{nllContribution}a as follows,
\be \label{a1}
A^{(1)} \equiv - \frac{1}{t} \, J_{(A) \, \alpha}^{(1) \, a} \, (p_A, 
q, \eta) \,
\left( N^{\alpha \beta}(q,\eta) + \frac{1}{t} \, v_{B}^{\alpha} \, 
v_A^{\mu} \, {\Pi}_{\mu \, \nu}(q,\eta)
\, v_B^{\nu} \, v_A^{\beta} \right) \, J_{(B) \, \beta}^{(1) \, a} \, 
(p_B, q, \eta),
\ee
where ${\Pi}_{\mu \, \nu}(q, \eta)$ stands for the one loop gluon self-energy.
We now put this contribution into the first factorized form, Eq. (\ref{fact1}), 
isolating the plus polarization for jet A, and the minus polarization for jet B.  
At NLL in the amplitude, we need 
the soft function $S^{(1,1)}$, Eq. (\ref{fact1}) with $n = m = 1$, to accuracy ${\cal O}(\alpha_s)$. 
Using the tulip-garden formalism described in Appendix 
\ref{tulipGarden}, the contribution to the first term on 
the right hand side of Eq. 
(\ref{a1}) is given by the subtractions shown in Fig. 
\ref{amplSubtract}.
In accordance with the notation adopted in Appendix 
\ref{tulipGarden}, the dashed lines indicate that we have made soft 
approximations
on gluons that are cut by them. A dashed line cutting a gluon 
attached to jet $A$($B$) means that the gluon is attached to the
corresponding jet through minus(plus) component of its polarization. 
Since $q^{\pm} = 0$ in the Regge limit,
Eq. (\ref{momentaDefinition}), we have $N^{\mu \pm}(q) = g^{\mu \pm}$.
This implies that the contributions between the diagrams in Fig. 
\ref{amplSubtract}c and in
Fig. \ref{amplSubtract}d as well as between the diagrams in Fig. 
\ref{amplSubtract}e and in Fig. \ref{amplSubtract}f cancel each other.
Therefore only the zeroth-order soft function diagram in Fig. \ref{amplSubtract}b 
survives in the factorized form, Eq.\ (\ref{fact1}).
\begin{figure} \center
\includegraphics*{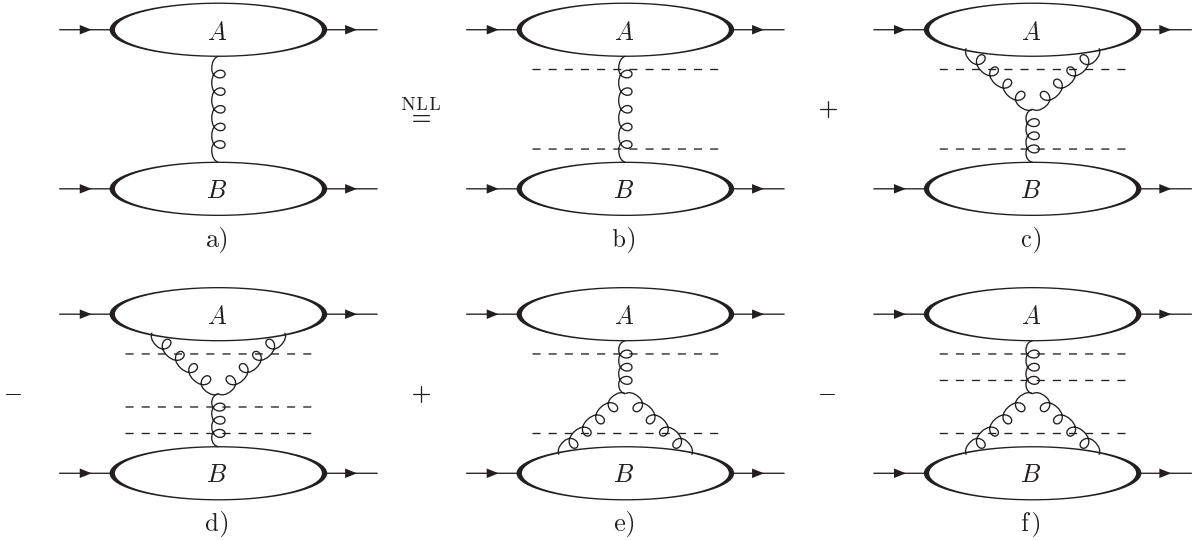}
\caption{\label{amplSubtract} Expansion of the one gluon exchange 
amplitude at NLL using the tulip garden formalism.}
\end{figure}

For the two gluon exchange, Fig.\ \ref{nllContribution}b, 
we only need the lowest order soft function at
NLL in the amplitude (and LL in singlet exchange).
The expression for the two gluon exchange diagram in Fig. 
\ref{nllContribution}b takes the form, Eq. (\ref{fact1}),
\be \label{a2}
A^{(2)} = \int \frac{{\rm d}^{D} k}{(2 \pi)^D} \, J_A^{(2) \; a \, b} 
(k^+ = 0, k^-, k_{\perp}) \, S(k^+, k^-, k_{\perp}) \,
J_B^{(2) \; a \, b} (k^- = 0, k^+, k_{\perp}),
\ee
where $S(k)$ is given by
\be \label{sk}
S(k) \equiv \frac{i}{2!} \, \frac{N^{- \, +}(k)}{k^2+i\epsilon} \, 
\frac{N^{- \, +}(q - k)}{(q-k)^2+i\epsilon}.
\ee
We have suppressed the dependence of the functions appearing in Eq. 
(\ref{a2}) on other arguments for brevity.
At NLL accuracy we are entitled to pick the plus Lorentz indices for 
jet function $J_A$ and the minus indices for jet function
$J_B$ only. We can also set $k^+ = 0$ in $J_A$ and $k^- = 0$ in $J_B$ 
since all the loop momenta inside the jets are collinear.
Eq. (\ref{a2}) represents the first factorized form, Eq. 
(\ref{fact1}), for the amplitude $A^{(2)}$.

Next, we follow the procedure described in Sec. 
\ref{secondFactorizedForm} to bring the amplitude into the second 
factorized form,
Eq. (\ref{fact2}).
We employ an identity based on Eq. (\ref{iden}), for the function 
$S(k)$ defined in Eq. (\ref{sk})
\bea \label{sIdentity}
S(k^+,k^-) & = & S(k^+ = 0, k^- = 0) \, \theta(M-|k^+|) \, 
\theta(M-|k^-|) \nonumber \\
& + & [S(k^+, k^- = 0) - S(k^+ = 0, k^- = 0) \, \theta(M-|k^+|) ] \, 
\theta(M-|k^-|) \nonumber \\
& + & [S(k^+=0, k^-) - S(k^+ = 0, k^- = 0) \, \theta(M-|k^-|)]  \, 
\theta(M-|k^+|) \nonumber \\
& + & [ \{S(k^+, k^-) - S(k^+, k^- = 0) \, \theta(M-|k^-|)\} - \nonumber \\
& & \{S(k^+ = 0, k^-) - S(k^+ = 0, k^- = 0) \, \theta(M-|k^-|)\} \, 
\theta(M-|k^+|)]. \nonumber \\
\eea
The contribution from the first term in Eq. (\ref{sIdentity}) gives 
immediately the second factorized form with $\Gamma^{(2)}_A$ and
$\Gamma^{(2)}_B$ defined in Eq. (\ref{gammaDef}) for $n=m=2$.

We now discuss the rest of the terms in Eq. (\ref{sIdentity}), which 
can be analyzed using the $K$-$G$ decomposition, since,
by construction, there is no contribution from the Glauber region. At 
the current accuracy only the $K$-gluon contributes.
After substituting the second term of Eq. (\ref{sIdentity}) into Eq. 
(\ref{a2}), we can factor the gluon with momentum $k$ from
jet $J_B^{(2)}$. However,
it is easy to verify, using the definitions for $K$ and $G$ gluons,
Eq. (\ref{kgDef}), the Ward identities, Eq.\ (\ref{gy2}), and the explicit components of 
the gluon propagator, Eq.\ (\ref{propagComp}), that
the $k^+$ integral is over an antisymmetric function.  As a result, 
this contribution vanishes.
In a similar fashion, the contribution from the third term in Eq. 
(\ref{sIdentity}), after used in Eq. (\ref{a2}), vanishes, since now
we can factor the soft gluon with momentum $k$ from jet $J_A^{(2)}$ 
and the $k^-$ integral is over an antisymmetric function.

In the case of the last term in Eq. (\ref{sIdentity}), after used in 
Eq. (\ref{a2}), we can factor the soft gluon with
momentum $k$ from both jets $J_A^{(2)}$ and $J_B^{(2)}$. The 
integrals of the soft function $S(k)$ over $k^+$ and $k^-$
are then
\be \label{sInteg}
\tilde{S}(k_{\perp}, q; M) \equiv  C_A \, \frac{g_s^2}{(2\pi)^2} \, 
\int_{-M}^{M} \frac{{\rm d} k^+}{k^+} \, \frac{{\rm d} k^-}{k^-}
\, S(k^+, k^-, k_{\perp}, q).
\ee
As usually, we leave the transverse momentum integral undone.
The $1/k^+$ and $1/k^-$ in the integral above are given by the 
Principal Value prescription because there is no contribution from
the Glauber region. Since the amplitude is independent on the choice 
of scale $M$, we can evaluate it
at arbitrary scale. We choose to work in the limit $M \rightarrow 0$. 
In this limit the contribution to the integral comes from the imaginary parts 
of the gluon propagators in Eq. (\ref{sk}), 
$-i \pi \delta(k^2)$ and $-i \pi \delta((k-q)^2)$. The integration 
is then trivial and Eq. (\ref{sInteg}) becomes
\be \label{sm0}
\tilde{S}(k_{\perp}, q) \equiv \lim_{M \rightarrow 0} 
\tilde{S}(k_{\perp}, q; M) \equiv  - C_A \, \frac{i g_s^2}{8}
\, \frac{1}{k_{\perp}^2 \, (k-q)^2_{\perp}}.
\ee
Combining the partial results of the analysis described above in Eq. 
(\ref{a2}),
we arrive at the second factorized form for the double gluon exchange 
amplitude,
Fig. \ref{nllContribution}b,
\bea \label{a2factored}
A^{(2)} & = & \int \frac{{\rm d}^{D-2} k_{\perp}}{(2 \pi)^{D-2}} \, 
{\Gamma}_A^{(2) \; a \, b} (k_{\perp}) \, \frac{1}{(2 \pi)^2} \,
S(k^+ = 0, k^- = 0, k_{\perp}) \,
{\Gamma}_B^{(2) \; a \, b} (k_{\perp}) \nonumber \\
& + & \int \frac{{\rm d}^{D-2} k_{\perp}}{(2 \pi)^{D-2}} \, 
{\Gamma}_A^{(1) \; a} (p_A, q) \; \tilde{S}(k_{\perp}, q) \;
{\Gamma}_B^{(1) \; a} (p_A, q).
\eea
Using Eq. (\ref{a1}) for $A^{(1)}$ and Eq. (\ref{a2factored}) for $A^{(2)}$,
we obtain the amplitude for the process (\ref{qqqq}) at NLL accuracy
\bea \label{ampNll}
A^{({\rm NLL})} & = & - \frac{1}{t} \, \Gamma_A^{(1) \, a} \, (p_A, q, \eta) \,
\left(1 + \frac{1}{t} \, {\Pi}_{+ \, -}(q,\eta) + \frac{i \pi}{2} \, 
\alpha^{(1)}(t) \right) \,
\Gamma_B^{(1) \, a} \, (p_B, q, \eta) \nonumber \\
& + & \int \frac{{\rm d}^{D-2} k_{\perp}}{(2 \pi)^{D-2}} \, 
{\Gamma}_A^{(2) \; a \, b} (k_{\perp}) \,
\frac{i}{8 \pi^2} \, \frac{1}{k_{\perp}^2 \, (k-q)^2_{\perp}} \, 
{\Gamma}_B^{(2) \; a \, b} (k_{\perp}).
\eea
In Eq. (\ref{ampNll}), we have used the explicit form for $S(k^+ = 0, 
k^- = 0, k_{\perp})$, which can be easily
identified from Eq. (\ref{sk}). We have also used the integral 
representation of the gluon trajectory given in Eq. (\ref{trajec1}).

In order to determine the high energy behavior of the amplitude in 
Eq. (\ref{ampNll}), we need to examine the high energy behavior
of $\Gamma_A^{(1)}$ or $\Gamma_B^{(1)}$ at NLL and the evolution of 
$\Gamma_A^{(2)}$ or $\Gamma_B^{(2)}$ at LL.
In this paper, we restrict the discussion of evolution equations to 
LL level, and hence we analyze the behavior of $\Gamma_A^{(2)}$ only.
We will address the study of NLL jet evolution, and gluon 
reggeization at this level, elsewhere \cite{tibor}.

We use the evolution equation given by Eq. (\ref{evolG}) in order to 
determine the LL dependence
of $\Gamma_A^{(2)}$ on $\ln(p_A^+)$.
In our special case of the two gluon exchange amplitude, it reads
\bea \label{evolGamma2}
\left(p_A^+ \, \frac{\partial}{\partial \, p_A^+} - 1 \right) 
{\Gamma}_A^{(2) \; a\, b} & = &
M \, \left[ J_A^{(2) \; a\, b}(k^- = +M, k^+ = 0, k_{\perp}) + 
J_A^{(2) \; a\, b}(k^- = -M, k^+ = 0, k_{\perp})\right] \nonumber \\
& - & {\tilde \eta}^{\alpha} \frac{\partial}{\partial \, 
{\eta}^{\alpha}} \, \Gamma_A^{(2) \; a\, b}.
\eea
\begin{figure} \center
\includegraphics*{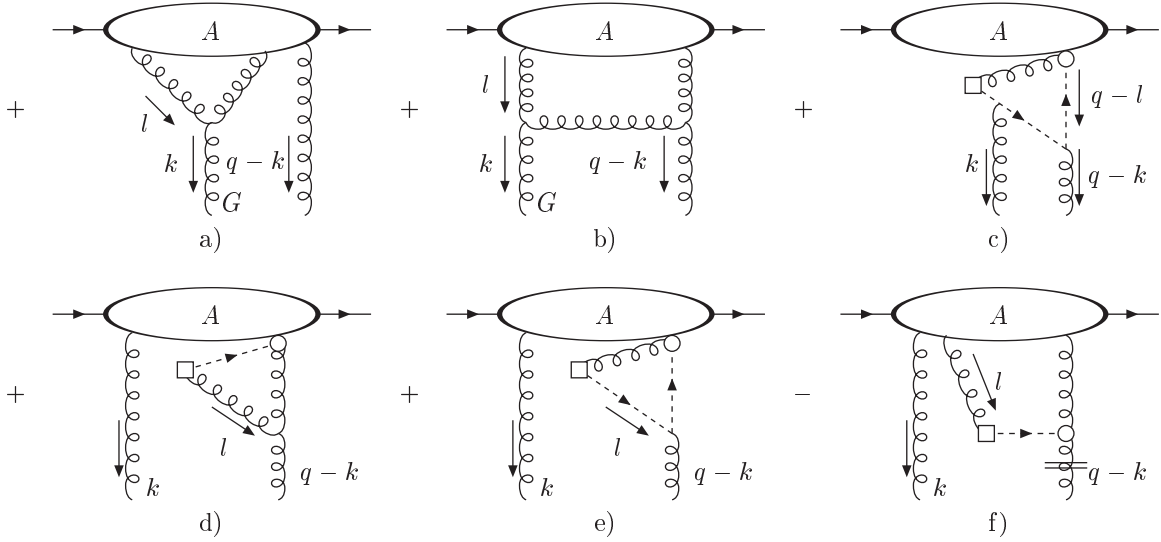}
\caption{\label{bfklDiagrams} Diagrams determining the evolution of 
$\Gamma_A^{(2)}$.}
\end{figure}
The first term in Eq. (\ref{evolGamma2}) can be analyzed using the 
$K$-$G$ decomposition. The contributions from the $K$-gluon cancel
between the $J_A^{(2)}(k^- = +M)$ and $J_A^{(2)}(k^- = -M)$. The 
contributions from the $G$ gluon, which we now discuss,
are shown in Figs. \ref{bfklDiagrams}a and \ref{bfklDiagrams}b.

Since the gluon with momentum $q-k$ in Fig. \ref{bfklDiagrams}a 
cannot be in the Glauber region, we can use $K$-$G$
decomposition on it. The $K$ part factors from $J_A^{(3)}$, while the 
$G$ part does not contribute at LL.
After factoring out the gluon
with momentum $q-k$ and performing the approximations on the jet 
function $J_A^{(2)}$, the contribution to Fig. \ref{bfklDiagrams}a
for $k^- = + M$ is
\be \label{fig10a}
{\rm Fig. \; \ref{bfklDiagrams}a} = -i g_s^2 f_{aec} f_{deb} \, 
\frac{1}{M} \int \frac{{\rm d}^D l}{(2 \pi)^D} \, S_1(k^+ = 0,
k^- = + M, k_{\perp}, l) \; J_A^{(2) \, c \, d}(l^+ = 0, l^-, l_{\perp}),
\ee
where we have defined
\be
S_1(k, l) \equiv \frac{N^{- \, \mu}(l)}{l^2} \, \frac{N^{- \, 
\nu}(k-l)}{(k-l)^2} \, V_{\mu \, \rho \, \nu}(l, -k, k-l) \,
\left( g^{\rho \, +} - \frac{k^{\rho}}{M} \right).
\ee
Next we follow the established procedure. First, we write
\be \label{s1Ident}
S_1(k,l) = S_1(k, l^- = 0) \, \theta(M-|l^-|) + \left[S_1(k,l) - 
S_1(k, l^- = 0) \, \theta(M-|l^-|)\right].
\ee
When we use the second term of Eq. (\ref{s1Ident}) in Eq. 
(\ref{fig10a}), we can factor the gluon with momentum $l$ from
$J_A^{(2)}$. Since the resulting integrand is an antisymmetric 
function under the simultaneous transformation $M \rightarrow -M$,
$l^{\pm} \rightarrow - l^{\pm}$, the contributions on the right hand 
side of Eq. (\ref{evolGamma2})
evaluated for $k^- = + M$ and $k^- = - M$ cancel each other.
Therefore we can write, using Eq. (\ref{s1Ident}) in Eq. (\ref{fig10a}),
\bea \label{fig10afact}
{\rm Fig. \; \ref{bfklDiagrams}a} & = & -i g_s^2 \, f_{aec} f_{deb} 
\, \frac{1}{M} \int \frac{{\rm d}^{D-2} l_{\perp}}{(2 \pi)^D}
\, \int {\rm d} l^+ \, S_1(k^+ = 0, k^- = + M, k_{\perp}, l^- = 0, 
l^+, l_{\perp}) \; {\Gamma}_A^{(2) \, c \, d}(l_{\perp})
\nonumber \\
& + & \ldots,
\eea
where by dots we mean the term which is canceled after we take into 
account the contributions to both $J_A^{(2)}(k^- = + M)$
and $J_A^{(2)}(k^- = - M)$ on the right hand side of Eq. (\ref{evolGamma2}).

Next, we perform the $l^+$ integral in Eq. (\ref{fig10afact}). As we 
have already mentioned above,
since the final result does not depend on the scale $M$, we can choose
arbitrary value of $M$. We have chosen to perform the calculation in 
the limit $M \rightarrow 0$.
Then the only nonvanishing contribution comes from the imaginary part 
of the propagator $1/[(l-k)^2 + i \epsilon]$,
$-i \pi \, \delta(2 M l^+ + (l-k)^2_{\perp})$. For this term the 
$l^+$ integration is trivial and we obtain
\be \label{fig10afactInteg}
M \, ( {\rm Fig. \; \ref{bfklDiagrams}a} ) = - \alpha_s \, f_{aec} 
f_{deb} \, \int \frac{{\rm d}^{D-2} l_{\perp}}{(2 \pi)^{D-2}}
\, \frac{2 k_{\perp} \cdot l_{\perp}}{l_{\perp}^2 \, (k-l)_{\perp}^2} 
\; {\Gamma}_A^{(2) \, c \, d}(l_{\perp}) + \ldots \, ,
\ee
which gives an $M$-independent contribution to the right hand side of 
Eq. (\ref{fig10a}).

We follow the same steps when dealing with the diagram in Fig. 
\ref{bfklDiagrams}b, whose soft subdiagram is given by
\be \label{s2}
S_2 (k, l) \equiv \frac{N^{- \, \mu}(l)}{l^2} \, \frac{N^{- \, 
\nu}(q-l)}{(q-l)^2} \, V_{\mu \, \rho \, \gamma}(l, -k, k-l) \,
\left( g^{\rho \, +} - \frac{k^{\rho}}{M} \right) \frac{N^{\gamma 
\delta}(l - k)}{(l-k)^2} \,
V_{\nu \, \delta \, -}(q - l, l - k, k - q).
\ee
First we use the identity (\ref{s1Ident}) for $S_2$.
The contribution due to the second term in Eq. (\ref{s1Ident}) 
vanishes, after the gluon with momentum $l$ has been factored
from $J_A^{(2)}$, due to the antisymmetry of the integrand.
Hence again, as in the case discussed above, only the term given by 
$S_2(l^- = 0, l^+, l_{\perp}, k)$ contributes.
In the limit $M \rightarrow 0$, the contribution comes from the 
imaginary part of the same denominator as in the case of Fig.
\ref{bfklDiagrams}a. The result is
\bea \label{fig10bfactInteg}
M \, ( {\rm Fig. \; \ref{bfklDiagrams}b} ) = & - & \alpha_s \, 
f_{aec} f_{deb} \, \int \frac{{\rm d}^{D-2} l_{\perp}}{(2 \pi)^{D-2}}
\, \frac{2}{l_{\perp}^2 \, (l-q)^2_{\perp} \, (k-l)_{\perp}^2} \nonumber \\
& \times & \left( k_{\perp}^2 l_{\perp}^2 - k_{\perp} \cdot l_{\perp} 
l_{\perp}^2 - k_{\perp} \cdot q_{\perp} l_{\perp}^2
- k_{\perp}^2 l_{\perp} \cdot q_{\perp} + 2 k_{\perp} \cdot l_{\perp} 
l_{\perp} \cdot q_{\perp} \right)  \times
{\Gamma}_A^{(2) \, c \, d}(l_{\perp}) \nonumber \\
& + & \ldots \, .
\eea
Combining the results of Eqs. (\ref{fig10afactInteg}) and 
(\ref{fig10bfactInteg}), we obtain the expression for the surface 
term in
Eq. (\ref{evolGamma2})
\bea \label{surfaceTerm}
\lefteqn{M \left[ J_A^{(2) \; a\, b}(k^- = +M, k^+ = 0, k_{\perp}) + 
J_A^{(2) \; a\, b}(k^- = -M, k^+ = 0, k_{\perp})\right]
= 2 \alpha_s \, f_{aec} f_{bed}} \nonumber \\
\hspace*{1cm} & \times & \int \frac{{\rm d}^{D-2} l_{\perp}}{(2 
\pi)^{D-2}} \left(
\frac{k_{\perp}^2}{l_{\perp}^2 \, (k-l)_{\perp}^2} + 
\frac{(k-q)_{\perp}^2}{(l-q)_{\perp}^2 \, (k-l)_{\perp}^2}
- \frac{q_{\perp}^2}{l_{\perp}^2 \, (q-l)_{\perp}^2} \right) \, 
\times {\Gamma}_A^{(2) \, c \, d}(l_{\perp}).
\eea

Next, we analyze the contributions to the term ${\tilde 
\eta}^{\alpha} \partial / \partial \, {\eta}^{\alpha} \, 
\Gamma_A^{(2)}$
in the evolution equation (\ref{evolGamma2}).
The contributing diagrams are shown in Figs. \ref{bfklDiagrams}c - 
\ref{bfklDiagrams}f. Note that for every diagram
in Figs. \ref{bfklDiagrams}c - \ref{bfklDiagrams}f, we have also 
diagrams when a loop containing the boxed vertex
is attached to the external gluon with momentum $k$, instead of to 
the external gluon with momentum
$q-k$.

In Fig. \ref{bfklDiagrams}c, we have to consider all the possible 
insertions of external gluons
with momenta $k$ and $q-k$. We have six possibilities. The 
contribution shown in Fig. \ref{bfklDiagrams}c is proportional to
(omitting the color factor)
\be \label{fig10c}
\int_{-M}^{M} {\rm d}k^- \, ({\rm Fig. \; \ref{bfklDiagrams}c}) 
\propto \int_{-M}^{M} {\rm d}k^- \,
\int \frac{{\rm d}^D l}{(2 \pi)^D} \frac{N^{- \, \mu}(l)}{l^2} 
\frac{S_{\mu}(l)}{l \cdot {\bar l}} \,
\frac{(\bar l - \bar k)^+}{(\bar l - \bar k)^2} \, \frac{(\bar q - 
\bar l)^+}{(\bar q - \bar l)^2} \, (k^+ = 0).
\ee
Since the integrand is an antisymmetric function under $k^- 
\rightarrow - k^-$ and $l^{\pm} \rightarrow - l^{\pm}$,
the integral in Eq. (\ref{fig10c}) vanishes. The same antisymmetry 
property holds for the remaining five diagrams and therefore,
there is no contribution from them.
\begin{figure} \center
\includegraphics*{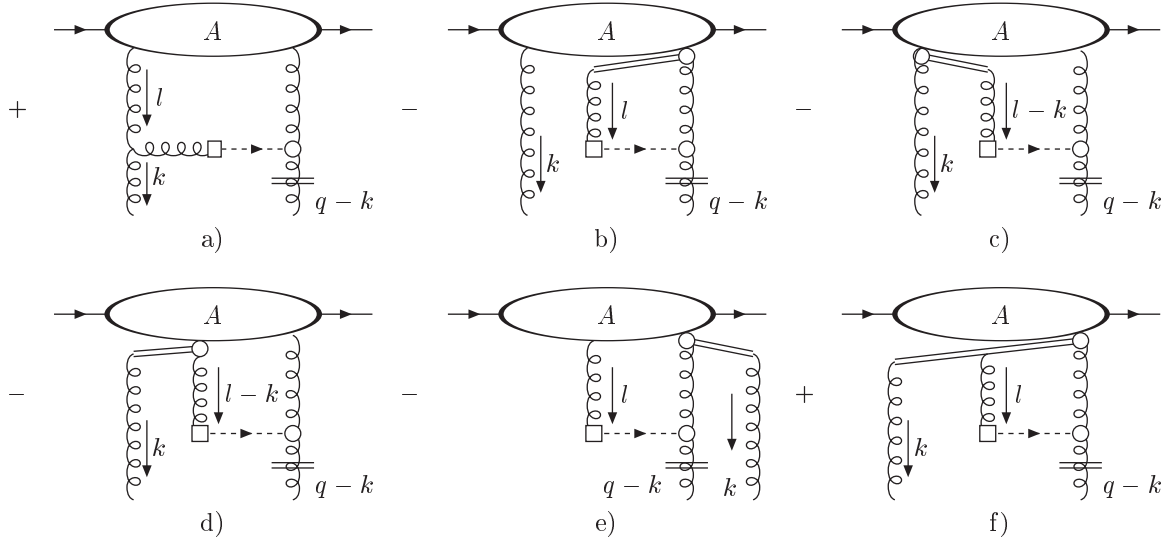}
\caption{\label{bfklFact} Contributions to the diagram in Fig. 
\ref{bfklDiagrams}f when the gluon coming out of the boxed vertex
is attached to the soft line (a) and when either or both gluons with 
momenta $k$ and $l$ are $K$ gluons and they are factored
from the jet (b - f).}
\end{figure}

Let us next focus on the diagram in Fig. \ref{bfklDiagrams}f.
When the gluon with momentum $l$ attaches to a soft line inside of 
the jet $J_A^{(3)}$, the contribution takes the form shown in
Fig. \ref{bfklFact}a.
If it attaches to a jet line, its contribution can be written as
\be \label{fig10f}
{\rm Fig. \; \ref{bfklDiagrams}f} = - g_s f_{b c d} \, \int 
\frac{{\rm d}^D l}{(2 \pi)^D} \, S_3(k^+ = 0,
k^-, k_{\perp}, l) \; J_A^{(3) \, a \, c \, d}(k^+ = 0, k^-, 
k_{\perp}, l^+ = 0, l^-, l_{\perp}),
\ee
with the soft function
\be \label{s3}
S_3(k,l) \equiv (q-k)^2 \, \frac{N^{- \mu}(l)}{l^2} \, 
\frac{S_{\mu}(l)}{l \cdot \bar l} \, \frac{N^{- +}(q-k-l)}{(q-k-l)^2}.
\ee
We use the identity for this soft function $S_3$, obtained from Eq. 
(\ref{sIdentity}) by the replacement $k^+ \rightarrow l^-$,
\bea \label{s3Ident}
S_3(l^-,k^-) & = & S_3(l^- = 0, k^- = 0) \, \theta(M-|l^-|) \, 
\theta(M-|k^-|) \nonumber \\
& + & [S_3(l^-, k^- = 0) - S_3(l^- = 0, k^- = 0) \, \theta(M-|l^-|) ] 
\, \theta(M-|k^-|) \nonumber \\
& + & [S_3(l^-=0, k^-) - S_3(l^- = 0, k^- = 0) \, \theta(M-|k^-|)] 
\, \theta(M-|l^-|) \nonumber \\
& + & [ \{S_3(l^-, k^-) - S_3(l^-, k^- = 0) \, \theta(M-|k^-|)\} \nonumber \\
& & - \{S_3(l^- = 0, k^-) - S_3(l^- = 0, k^- = 0) \, 
\theta(M-|k^-|)\} \, \theta(M-|l^-|)],
\eea
to treat the soft gluons with momenta $k$ and $l$ attached to jet $J_A^{(3)}$.
The contribution from the first term in Eq. (\ref{s3Ident}), when 
used in Eq. (\ref{fig10f}), vanishes since the integrand
$S_3(k^+=k^- = 0, k_{\perp}, l^- = 0, l^+, l_{\perp})$ is an 
antisymmetric function of $l^+$, as can be easily checked
using Eqs. (\ref{propagComp}), (\ref{sCompon}) and (\ref{s3}).
We can apply the $K$-$G$ decomposition on the gluon with momentum $l$ 
when treating the
second term in Eq. (\ref{s3Ident}) used in Eq. (\ref{fig10f}). At LL 
only the $K$ gluon contributes. It can be factored from
the jet function $J^{(3)}_A$ with the result shown in Figs. 
\ref{bfklFact}b and \ref{bfklFact}c.
In a similar way we can treat the gluon with momentum $k$ in the 
third term of Eq. (\ref{s3Ident}).
After we factor this gluon from the jet $J_A^{(3)}$, we obtain the 
contributions shown in Figs. \ref{bfklFact}d and
\ref{bfklFact}e. In the case of the last term in Eq. (\ref{s3Ident}), 
we can factor out both soft gluons with momenta
$k$ and $l$ from jet $J_A^{(3)}$. The result of this factorization is 
shown in Fig. \ref{bfklFact}f.

Next, we note that the combination of the diagrams in Figs. 
\ref{bfklDiagrams}d, \ref{bfklDiagrams}e and \ref{bfklFact}b is the 
same
as the result encountered in the analysis of the LL amplitude, Fig. 
\ref{trajll}. We write
\be \label{bfklTraj2}
\int_{-M}^{M} {\rm d} k^- \left({\rm Fig. \; \ref{bfklDiagrams}d} + 
{\rm Fig. \; \ref{bfklDiagrams}e} + {\rm Fig. \; \ref{bfklFact}b}
\right) = \alpha^{(1)}(q_{\perp}-k_{\perp}) \, \Gamma_A^{(2) \; a \, 
b} (p_A, q, k_{\perp}).
\ee
where $\alpha^{(1)}(q-k)$ in Eq. (\ref{bfklTraj2}) is given by the 
diagrams in Fig. \ref{trajll} with an external momentum
$q-k = (0^+, 0^-, q_{\perp} - k_{\perp})$.
In the case when the gluon coming out of the boxed vertex attaches to 
an external gluon with momentum $k$, we evaluate
the one loop trajectory $\alpha^{(1)}(k_{\perp})$ in Eq. (\ref{bfklTraj2}).

To complete the analysis, we have to discuss the diagrams in Figs. 
\ref{bfklFact}a and \ref{bfklFact}c - \ref{bfklFact}f.
In the region $l^{\pm} \sim l_{\perp}$, we can factor the gluon with 
momentum $l$ from the jet function
$J_A^{(2)}(l^+ = 0, l^-, l_{\perp})$ in the case of the diagram in 
Fig. \ref{bfklFact}a.
The resulting $k^-$ and $l^{\pm}$ integral is over an antisymmetric 
function of $k^-$ and $l^{\pm}$, and therefore
it vanishes. So the only contribution comes from the Glauber region, 
where we can set $l^- = 0$ outside
$J_A^{(2)}(l^+ = 0, l^-, l_{\perp})$. As above, we perform the $l^+$ 
and $k^-$ integrals in the limit $M \rightarrow 0$.
The integrand does not develop a singularity in $k^-$ and/or $l^+$ 
strong enough to compensate
for the shrinkage of the integration region $\int_{-M}^{M} {\rm 
d}k^-$ when $M \rightarrow 0$. Hence the diagram
in Fig. \ref{bfklFact}a does not contribute in the limit $M 
\rightarrow 0$. In a similar way as for the diagram in
Fig. \ref{bfklFact}a, none of the diagrams in Figs. \ref{bfklFact}c - 
\ref{bfklFact}f contribute. The diagrams
in Figs. \ref{bfklFact}c - \ref{bfklFact}e vanish in the $M 
\rightarrow 0$ limit, while in the case of the diagram
in Fig. \ref{bfklFact}f the $k^-$ and $l^{\pm}$ integral is over an 
antisymmetric function of $k^-$ and $l^{\pm}$.

At this point we have discussed all the contributions appearing on 
the right hand side of the evolution equation (\ref{evolGamma2}).
Combining the partial results given by Eqs. (\ref{bfklTraj2}) and 
(\ref{surfaceTerm}) in Eq. (\ref{evolGamma2}),
we arrive at the evolution equation governing the high energy 
behavior of $\Gamma_A^{(2)}$
\bea \label{evolGamma2Result}
\left(p_A^+ \, \frac{\partial}{\partial \, p_A^+} - 
1\right){\Gamma}_A^{(2) \; a\, b} (p_A^+, q, k_{\perp}) & = &
2 \alpha_s \, f_{aec} f_{bed} \, \int \frac{{\rm d}^{D-2} 
l_{\perp}}{(2 \pi)^{D-2}} \;
\Gamma_A^{(2) \; c\, d} (p_A^+, q, l_{\perp}) \nonumber \\
& \times & \left( \frac{k_{\perp}^2}{l_{\perp}^2 \, (k-l)_{\perp}^2} 
+ \frac{(k-q)_{\perp}^2}{(l-q)_{\perp}^2 \, (k-l)_{\perp}^2}
- \frac{q_{\perp}^2}{l_{\perp}^2 \, (q-l)_{\perp}^2} \right) \nonumber \\
& + & \left( \alpha^{(1)}(k_{\perp}) + 
\alpha^{(1)}(q_{\perp}-k_{\perp}) \right) \times \Gamma_A^{(2) \; a\, 
b} (p_A^+, q, k_{\perp}).
\eea
Projecting out onto the color singlet in Eq. 
(\ref{evolGamma2Result}), we immediately recover the celebrated BFKL 
equation
\cite{bfkl}.

\subsection{Evolution of $\Gamma^{(n)}$ at LL} \label{gamman}

We can now generalize Eq. (\ref{evolGamma2Result}) to the case of 
$\Gamma_A^{(n)}$.
The evolution kernel in this case contains, besides a piece diagonal 
in the number of external gluons,
also contributions which relate jet functions with different number 
of external gluons
\bea \label{evolGammaN}
\lefteqn{\left(p_A^+ \, \frac{\partial}{\partial \, p_A^+} - 1\right) 
{\Gamma}_A^{(n) \; a_1 \ldots \, a_n} (p_A^+, q, k_{1 \, \perp},
  \dots, k_{n \, \perp}) = } \nonumber \\
& & 2 \alpha_s \, \sum_{i<j}^{n} f_{a_i \, e \, b_i} f_{a_j \, e \, 
b_j} \, \int \frac{{\rm d}^{D-2} l_{i \, \perp}}{(2 \pi)^{D-2}} \,
\frac{{\rm d}^{D-2} l_{j \, \perp}}{(2 \pi)^{D-2}} \, 
\delta^{(2)}(l_{i \, \perp} + l_{j \, \perp} - k_{i \, \perp} - k_{j 
\,
\perp}) \nonumber \\
& & \times \; \left( \frac{k_{i \, \perp}^2}{l_{i \, \perp}^2 \, 
(k_i-l_i)_{\perp}^2} + \frac{k_{j \, \perp}^2}{l_{j \, \perp}^2
\, (k_j - l_j)_{\perp}^2} - \frac{(k_i + k_j)_{\perp}^2}{l_{i \, 
\perp}^2 \, l_{j \, \perp}^2} \right) \nonumber \\
& & \times \; \Gamma_A^{(n) \; a_1 \ldots \, b_i \ldots \, b_j \ldots 
\, a_n} (p_A^+, q, k_{1 \, \perp}, \ldots,
l_{i \, \perp}, \ldots, l_{j \, \perp}, \ldots, k_{n \, \perp}) \nonumber \\
& & + \; \sum_{i = 1}^{n} \left( \alpha^{(1)}(k_{i \, \perp}) \right) \times
{\Gamma}_A^{(n) \; a_1 \ldots \, a_n} (p_A^+, q, k_{1 \, \perp}, 
\dots, k_{n \, \perp}) \nonumber \\
& & + \; \sum_{n' = 1}^{n-1} {\cal K}^{(n,n')}_{a_1 \ldots \, a_n; \; 
b_1 \ldots \, b_{n'}} \, {\otimes}_{\perp}
\, {\Gamma}_A^{(n')\; b_1 \ldots \, b_{n'}},
\eea
where ${\otimes}_{\perp}$ denotes a convolution in transverse 
momentum space. The last term in Eq. (\ref{evolGammaN}) corresponds 
to the
configurations when one or more external gluons attach to a gluon or 
a ghost lines forming the one loop kernel derived
for $\Gamma_A^{(2)}$.
Using the notation of Sec. \ref{solEvolEq}, we can write Eq. 
(\ref{evolGammaN}) at $r$-loop order in a form
\be \label{evolGamman}
\left(p_A^+ \frac{\partial}{\partial p_A^+} - 1 \right) \Gamma_A^{(n,r)}
= \sum_{n'=1}^{n} {\cal K}^{(n,n';1)} \otimes \Gamma_A^{(n',r-1)}.
\ee
It corresponds to Eq. (\ref{coeff2}) of Sec. \ref{solEvolEq} when written
in terms of the coefficients $c_r^{(n,r)}$
introduced in Eq. (\ref{gammaExpand}).
From Eq. (\ref{evolGamman}) we immediately see that the following 
property of the one loop kernel
\be \label{ktheta}
{\cal K}^{(n,n';1)} = \theta(n-n') \, {\tilde {\cal K}}^{(n,n';1)},
\ee
is satisfied. We recall that this step was essential in demonstrating 
that the set of evolution equations, Eq. (\ref{evolG}),
forms a consistent system, refer to the paragraph above Eq. (\ref{coeff2}).

The term diagonal in the number of external gluons in Eq. 
(\ref{evolGammaN}) coincides
with the evolution equation derived in Ref. \cite{jaroszewicz}.
Our formalism, besides enabling us to go systematically beyond LL 
accuracy, Ref. \cite{tibor},
indicates that even at LL, in addition to the kernels found in Ref.\ 
\cite{jaroszewicz},
the kernel has contributions which relate jet functions with different 
number of external gluons.

\section{Conclusions}

We have established a systematic method that shows that
it is possible to resum the
large logarithms appearing in the perturbation series
of scattering amplitudes for $ 2 \rightarrow 2 $ partonic processes
to arbitrary logarithmic accuracy in the Regge limit.
Up to corrections suppressed by powers of $|t|/s$, the amplitude can
be expressed as a sum of convolutions in transverse momentum space
over soft and jet functions, Eq. (\ref{fact2}). All the large
logarithms are organized in the jet functions, Eq. (\ref{gammaDef}).
They are resummed using Eqs. (\ref{evolG}) and/or (\ref{evol}).
The evolution kernel $\cal K$ in Eq. (\ref{evol}) is a calculable
function of its arguments order by order in perturbation theory.
This is the central result of our analysis.

As an illustration of the general algorithm we have demonstrated it
in an action at NLL
for the amplitude and LL for the evolution equations. We reserve the
study of the NLL evolution,
which addresses the reggeization of a gluon at NLL, for future work
\cite{tibor}.

The derivation of the evolution equations and the procedure for
finding the kernels
was given above in Coulomb
gauge.  Clearly, it will be useful and interesting to reformulate our
arguments in covariant gauges.  
In addition, the connection of our formalism to the 
effective action approach to small-$x$ and the Regge limit, 
Refs. \cite{smallx,balitski} should provide further insight.

\subsection*{Acknowledgment}

I wish to thank G. Sterman for suggesting this problem to me, for 
invaluable help and constant encouragement.
I also wish to thank A. Sen for very helpful discussions. During the 
process of this work I have benefited from conversations
with C.F. Berger, G.T. Bodwin, J.C. Collins, P.A. Grassi, M.E. 
Tejeda-Yeomans, A.R. White and K. Zoubos.

\appendix

\section{Power counting with contracted vertices} \label{contracted vertices}

In this appendix we will include the possibility of contracted 
vertices in the reduced diagram
in Fig. \ref{reduced}a. These are associated with internal lines 
(collapsed to a point) which are off-shell by
$\sqrt{s}$. Our analysis closely follows \cite{sterman78} and \cite{sen81}.

If we go back to the argument that led us to Eq. (\ref{os}) for the 
superficial degree of IR
divergence for the soft part, we see that the same reasoning as in 
the case of elementary vertices applies to the case of
contracted vertices since the result (\ref{os}) has been obtained by 
means of dimensional counting.

The analysis of contracted vertices connecting jet lines only is, 
however,  more subtle.
We have to demonstrate that the suppression factors corresponding to 
the contracted vertices are at least as great as
the ones for the elementary vertices. The expression (\ref{oa3}) 
tells us that we can restrict ourselves to the two and
three point vertices.  For these cases, we analyze the full two and 
three-point subdiagrams, by
studying the tensor structures that are found after integration over 
their internal loop momenta.

Before we discuss all the possible structures, we state some results 
which will be essential for the upcoming analysis.
The first one is the simple Dirac matrix identity
\be \label{slash}
\not{\!a} \, \not{\!b} \, \not{\!a} = 2 (a \cdot b) \, \not{\!a} - 
a^2 \, \not{\!b}.
\ee
The other two follow from Eqs. (\ref{propagator}) and 
(\ref{propagComp}) for the gluon propagator in Coulomb gauge,
and hold for any jet momenta scaling as
$l_A \, \sim \, l_A^{\prime} \, \sim \, 
\sqrt{s}(1^+,\lambda^-,{\lambda}^{1/2})$ collinear to the
momentum $p_A$ defined in Eq. (\ref{momentaDefinition})
\bea \label{contractPropagator}
l_A^{\prime \; \alpha} \, N_{\alpha \beta}(l_A, \eta) & = & {\cal 
O}({\lambda}^{1/2} \, \sqrt{s}), \nonumber \\
{\bar l}_A^{\prime \; \alpha} \, N_{\alpha \beta}(l_A, \eta) & = & 
{\cal O}({\lambda}^{1/2} \, \sqrt{s}),
\eea
for all components of $\beta$.
We now proceed to discuss the particular cases.

{\it Ghost self-energy:} The most general covariant structure is, 
using $p \cdot \bar p = {\bar p}^2$,
\footnote {In the rest of this subsection we are concerned the 
momentum factors only, and we omit dependence on the color structure.}
\be \label{ghostSelfEnergy}
\Pi(p, \bar p) = p \cdot {\bar p} \, f(p^2 / {\mu}^2, \, {\bar p}^2 / 
{\mu}^2, \, {\alpha}_s(\mu)),
\ee
where $\mu$ is a scale introduced by a UV/IR regularization of 
Feynman diagrams and $p$ is the momentum of an internal jet line.
Strictly speaking, the covariants should be formed
from the vectors $p$ and $\eta$, but since $p$ has nonzero light-cone 
components, we can use
Eq. (\ref{barVector}), to express $\eta$
in terms of $\bar p$.
The maximum degree of divergence for the ghost self-energy occurs 
when the internal lines become either parallel to the external 
momentum $p$ or soft. The most general pinch singular
surface consists of a subdiagram of collinear lines moving in a 
direction of the external ghost.
This subdiagram can interact with itself through the exchange of soft quanta.
Power counting arguments similar to the ones given in Sec. 
\ref{elementary vertices} show, however, that there is no IR 
divergence for these pinch singular points. This shows that the 
dimensionless function $f$ in Eq. (\ref{ghostSelfEnergy}) is IR 
finite. Hence the combination [tree level ghost propagator] - [ghost 
self-energy] - [tree level ghost propagator], $[1/(p \cdot {\bar p})] 
\, \Pi (p, \bar p) \,
[1/(p \cdot {\bar p})]$, is suppressed at least as much as a single 
tree level ghost
propagator, $1/(p \cdot {\bar p})$. Therefore the contracted two 
point ghost vertex within a jet subdiagram
contributes at least the same suppression as a single tree level 
ghost propagator.

{\it Gluon self-energy:} With external momentum $p$, its most general 
tensor decomposition has the form
\be \label{gluonSelfEnergy}
{\Pi}_{\mu\nu}(p, \bar p) = g_{\mu\nu} \, p^2 \, f_1 + p_{\mu}p_{\nu} 
\, f_2 + {\bar p}_{\mu}{\bar p}_{\nu} \, f_3 +
(p_{\mu}{\bar p}_{\nu} + {\bar p}_{\mu}p_{\nu}) \, f_4 \, .
\ee
As verified by explicit one-loop calculations in Refs. 
\cite{leibbrandt96} and \cite{leibbrandt98}
the gluon self-energy in Coulomb gauge is not transverse.
In Eq. (\ref{gluonSelfEnergy}), the $f_i = f_i \, ( p^2 / {\mu}^2, \, 
{\bar p}^2 / {\mu}^2, \, {\alpha}_s(\mu))$
are dimensionless functions.
Contracting ${\Pi}_{\mu\nu}$ with tree level gluon propagators, and 
using Eq. (\ref{gluonProperties}),
the last two terms in Eq. (\ref{gluonSelfEnergy}) drop out and the 
first and the second terms give at least one factor of
$p^2$ in the numerator, which cancels one  of the $(1/p^2)$ 
denominator factors.
Since the maximum degree of IR divergence for the gluon self-energy 
occurs when all the internal lines
become either collinear to the external momentum $p$ or soft, we can 
use the results of the power counting
of Sec. \ref{elementary vertices} to demonstrate that the 
dimensionless functions $f_i$ are at
worst logarithmically divergent.  Therefore the combination: gluon 
jet line - 2 point gluon contracted vertex - gluon jet
line, behaves the same way as a gluon jet line for the purpose of the 
jet power counting.

{\it Fermion self-energy:} In the massless fermion limit, the most 
general matrix structure of the
fermion self-energy is
\be \label{sigma}
\Sigma(p,{\bar p}) = \not{\!p} \, g_1 + \not{\!{\bar p}} \, g_2,
\ee
with dimensionless functions $g_{i}=g_i(p^2/{\mu}^2, \, {\bar 
p}^2/{\mu}^2, \, {\alpha}_s(\mu)), \; i=1,2$.
When we sandwich the fermion self-energy between the tree level 
fermion denominators, the first term in Eq. (\ref{sigma}) behaves
the same way as the tree level fermion  propagator, modulo 
logarithmic enhancements due to the function $g_1$.
The second term, however, is absent from the fermion self-energy as 
was shown in Ref. \cite{sen81} using the method of
induction and Ward identities. The idea was to study a variation of 
the fermion self-energy by making an infinitesimal
Lorentz boost on the external momentum. This implies a relationship 
between the $(r+1)$ and the $r$-loop self
energy. Assuming that the term proportional to $\not{\!\!{\bar p}}$ 
is absent from the $r$-loop expansion Sen shows that
it is also absent from the $(r+1)$-loop expansion. So the first term 
in Eq. (\ref{sigma}) is the only possible structure
of the fermion self-energy when its external momentum is jet like and 
approaches mass shell.

Now let us investigate the 3 point functions. \\
{\it Fermion-gluon-fermion vertex function:} ${\Gamma}_{\mu}$, can 
depend on vectors that scale as $l_A, \, l_A^{\prime}$
in Eq. (\ref{contractPropagator}), provided all momenta external to 
the contracted vertex are collinear to momentum $p_A$
given in Eq. (\ref{momentaDefinition}). It has one Lorentz index, 
$\mu$, and neglecting the fermion masses, it contains an odd number 
of gamma matrices. This implies that the most general tensor and 
gamma matrix expansion of ${\Gamma}_{\mu}$ involves
\begin{enumerate}
\item ${\gamma}_{\mu}$,
\item ${\gamma}_{\mu}\not{\!l_A}\not {\!{\bar l}_A} \, / \, (l_A 
\cdot {\bar l}_A)$ and all permutations of ${\gamma}_{\mu}, \, 
\not{\!l_A}, \, \not {\!{\bar l}_A}$,
\item  $\not{\!l_A} \, l_A^{\mu} \, / \, l_A^2$, \, $\not{\!{\bar 
l}_A} \, l_A^{\mu} \, / \, ({\bar l}_A \cdot l_A)$, \,
$\not{\!l_A} \, {\bar l}_A^{\mu} \, / \, (l_A \cdot {\bar l}_A)$, \, 
$\not{\!{\bar l}_A} \, {\bar l}_A^{\mu} \, / \, {{\bar l}_A}^2$.
\end{enumerate}
The differences between the listed set of structures and other 
possible combinations are ${\cal O}({\lambda}^{1/2} \, \sqrt{s})$, as
  can be shown using Eqs. (\ref{slash})-(\ref{contractPropagator}).
The listed gamma matrix structures are multiplied by dimensionless 
functions, which can depend on the combinations
$l_A^2, \; {\bar l}_A^2, \; l_A^{\prime \; 2}, \; {\bar l}_A^{\prime 
\; 2}$, besides the renormalization scale
and the running coupling. Using the arguments similar to the ones 
leading to Eq. (\ref{oa4}),
we easily verify that the above mentioned dimensionless functions 
are at most logarithmically divergent.
Next we analyze the possible Dirac structures.
\begin{enumerate}
\item The first case has the same structure as the elementary vertex, 
and therefore causes the same suppression as the
elementary vertex.
\item The fermion-gluon-fermion composite 3-point vertex is 
sandwiched between the factors $\not{\!l_A^{\prime}}$ and 
$\not{\!l_A}$, originating from the numerators of the fermion 
propagators external to the composite vertex. Therefore the terms 
from case 2 where $\not{\!l_A}$ is on the first or third position in 
the string of the gamma matrices provide a suppression 
$\sqrt{l_A^2}$. On the other hand in the case, when $\not{\!l_A}$ is 
in the middle of this string of three gamma matrices, we encounter 
the combination
$\not{\!l_A^{\prime}} \, {\gamma}_{\mu} \not{\!l_A}$ after taking 
into account the numerators of the external fermions.
Using Eq. (\ref{slash}), we can immediately recognize that this 
combination provides a suppression ${\lambda}^{1/2}$.
\item Based on the preceding arguments it is obvious that also the 
structures included in item 3 supply at least the same suppression
factor as the elementary vertex.
\end{enumerate}
Therefore, we conclude that the composite 3-point 
fermion-gluon-fermion vertex behaves as the elementary vertex for the 
purposes
of the jet power counting.

{\it Three gluon vertex:} $V_{\mu\nu\rho}$, with external momenta 
collinear to momentum $p_A$.
This vertex can depend on momenta
$l_A, \, {\bar l}_A$ defined above and the metric tensor $g_{\alpha\beta}$.
Taking into account the dimension of the 3 gluon Green function, its 
only possible tensor structure involves combinations
of the form $[g_{\mu \nu} \, l_A^{\rho} + \mathrm{perm.} + {\cal 
O}({\lambda}^{1/2} \, \sqrt{s})]$ and
$[l_A^{\mu} \, l_A^{\nu} \, l_A^{\rho} / l_A^2 + {\cal 
O}({\lambda}^{1/2} \, \sqrt{s})]$,
with all possible replacements of $l_A \rightarrow {\bar l}_A$. These 
tensor structures are multiplied by dimensionless functions.
The former is the same as in the case of an elementary vertex and it 
therefore supplies the same suppression factor
as the elementary vertex. The latter also provides the same 
suppression as the elementary vertex, since the two momenta,
say $l_A^{\mu}, \, l_A^{\nu}$, after being contracted with the 
propagators of the external gluons, give suppression factors,
as in Eq. (\ref{contractPropagator}), which cancel the $1/l_A^2$ 
enhancement. The leftover momentum $l_A^{\rho}$
provides the same suppression factor as the elementary vertex.
Using the collinear power counting of Sec. \ref{elementary vertices}, 
one can immediately see that the
IR divergence of the dimensionless functions multiplying these tensor 
structures is not worse than logarithmic.
Hence, there is a suppression factor ${\lambda}^{1/2}$ associated 
with every contracted 3 gluon vertex.

{\it Ghost - gluon - antighost three point vertex:} When all lines 
external to the contracted vertex
are of the order $l_A$, the most general tensor structure for this 
contracted vertex is
\be \label{ghostVertex}
l_A^{\mu} \, h_1 + {\bar l}_A^{\mu} \, h_2 + {\cal O}({\lambda}^{1/2} 
\, \sqrt{s}),
\ee
with dimensionless functions $h_i = h_i \, ( l_A^2 / {\mu}^2, \, 
{\bar l}_A^2 / {\mu}^2, \, {\alpha}_s(\mu))$, $i = 1, 2$,
which are at most logarithmically IR divergent. Using Eq. 
(\ref{contractPropagator}), we see that when the momenta in Eq.
(\ref{ghostVertex}) are contracted with the tree level gluon 
propagator, we get a suppression of the order of the transverse
jet momentum, and that this contracted vertex gives the same 
suppression as the elementary three point vertex, at least.

\section{Varying the Gauge-Fixing Vector} \label{variation}

In this appendix we study the effect of an infinitesimal boost, 
performed on the gauge fixing vector $\eta$, on an expectation of a
time ordered product of fields, denoted by $O$, taken between physical states.
The gauge-fixing and the ghost terms in the QCD lagrangian are
\bea \label{lagrangian}
{\cal L}_{\mathrm{g.f.}} (x) & = & - \frac{1}{2 \xi} \, g_a^2 (x), \nonumber \\
{\cal L}_{\mathrm{ghost}} (x) & = & - b_a(x) \, \delta_{\mathrm{BRS}} 
\, g_a(x) \, / \delta \Lambda,
\eea
respectively. In Eq. (\ref{lagrangian}), $\delta \Lambda$ is a 
Grassmann parameter defining the BRS transformation, $b_a(x)$ is
an antighost field and
\be
g_a(x) \equiv - {\bar \partial} \cdot A_a (x) \equiv - (\partial - 
(\eta \cdot \partial) \, \eta) \cdot A_a(x).
\ee
Let us consider an infinitesimal boost with velocity $\delta \beta$ 
on a gauge fixing vector $\eta $ performed in the plus-minus plane
\be
\eta \rightarrow \eta' \equiv \eta + {\tilde \eta} \, \delta \beta,
\ee
where the vectors $\eta$ and ${\tilde \eta}$ are defined in Eqs. 
(\ref{eta}) and (\ref{etaTilde}), respectively.
Since only the gauge fixing and the ghost terms in the QCD lagrangian 
depend on $\eta$, we can write to accuracy
${\cal O}(\delta \beta ^2)$
\bea
\delta <{O}> \, & \equiv & \, <{O}(\eta')> - <{O}(\eta)> \; = \; 
<{\tilde \eta}^{\alpha} \frac{\partial \, {O}}
{\partial {\eta}^{\alpha}} \, \delta \beta> \nonumber \\
& = & - \frac{i}{\xi} \, \int {\mathrm d}^4 x <{O}
(\eta) \, g_a (x) \, \delta g_a (x)> - i \int {\mathrm d}^4 x < {O} 
(\eta) \, b_a(x) \, \delta \, (\delta_{\mathrm{BRS}} \, g_a (x) / \delta 
\Lambda)>. \nonumber \\
& &
\eea
Using the BRS invariance of the QCD lagrangian and the BRS 
transformation rule for an antighost field
\be
\delta_{\mathrm{BRS}} b_a(x) / \delta \Lambda \, = \, \frac{1}{\xi} \, g_a(x),
\ee
we arrive at
\be \label{var}
\delta <{O}> = -i \int {\mathrm d}^4 x <(\delta_{\mathrm{BRS}} {O} / 
\delta \Lambda) \, b_a(x) \, \delta \, g_a(x)>.
\ee
Taking a variation of $g_a(x)$ in Eq. (\ref{var}), we obtain
\be \label{varEquation}
\delta <{O}> = -i \int {\mathrm d}^4 x <(\delta_{\mathrm{BRS}} \, {O} 
/ \delta \Lambda) \, b_a(x) \,
(({\tilde \eta} \cdot \partial) \, \eta + (\eta \cdot \partial) \, 
{\tilde \eta}) \cdot A_a(x)>.
\ee
Substituting for $O$ a product of $n$ gluon fields, we can use Eq. 
(\ref{varEquation}), together with the rule
for the BRS transformation of a gluon field
\be \label{aBRS}
\delta_{\mathrm{BRS}} \, A_{\mu}^a(x) / \delta \Lambda = 
\partial_{\mu} c^a(x) + g_s f^{abc}A_{\mu}^b(x)c^c(x),
\ee
with $c^a(x)$ representing the ghost field, to derive the gauge 
variation for a connected
Green function.   However, our jet functions are one-particle irreducible
in external soft lines and we therefore cannot apply Eq. 
(\ref{varEquation}) directly, and must find an analog for
this subset of diagrams.
The modification of Eq.\ (\ref{varEquation}) due to the restriction 
to 1PI diagrams is, however, not difficult to identify.

Let us consider the graphical analog of the derivation
of Eq.\ (\ref{varEquation}) just outlined.  The variation in
$\eta$ may be implemented as a change in the gluon propagator and, in
Coulomb gauge, the ghost-gluon
interaction, which is also $\eta$-dependent.  This is the viewpoint
that was taken in axial gauge
in Ref.\ \cite{coso}.
At lowest order in the variation,
the modified gluon propagator produces scalar-polarized gluon lines,
which decouple through
repeated applications
of tree-level Ward identities to the sum over all diagrams.  The
relevant tree-level identities
are given in \cite{thooft}.
We need not describe these identities in
detail here.  We need only note that they are to be applied to any
diagram in which
a scalar polarized gluon appears at an internal vertex.  Every such application
produces a sum of diagrams, each of which
fall into one of two sets: 1)  diagrams in which an internal gluon
line is transformed
to a yet another ghost line ending in a scalar polarization, and 2)
diagrams in
which one gluon line is contracted to a point.  The new vertex
formed in the former case is the ghost term, and in the
latter case it is the ghost-gluon vertex of the BRS variation
(\ref{aBRS}).
Eq. (\ref{varEquation}) must result from the cancellation of all diagrams,
set 2), in which an internal gluon line is contracted.  Contracted external
lines provide the ghost-gluon terms,
and the ghost lines of set 1) eventually provide the ghost terms of 
the BRS variations (\ref{aBRS})
of external fields in Eq. (\ref{varEquation}).

The simplicity of the tree level Ward identities
puts strong limitations on the sets of diagrams that can combine
to form different diagrammatic contributions to Eq. (\ref{varEquation}).
For diagrams of set 1), the topology of the original diagram is
unchanged, and a 1PI diagram remains
1PI.  For diagrams of set 2), generally 1PI diagrams remain 1PI, except in the
special case of a diagram that is two-particle reducible, with these
two lines separated by a single propagator.
In this case, the contraction of the
internal line that separates the other two will bring those two lines together
at a single vertex, producing a diagram precisely of the topology
shown in Fig.\
\ref{variationJ}.   On the one hand, by Eq.\ (\ref{varEquation}) all
such diagrams must cancel
in the full perturbative sum. On the
other hand, the same topology results from a diagram that is
one-particle {\it reducible}
with respect to a single line, which is then contracted as a result of the
tree-level Ward identity.
The latter diagram, however, is not included in
the set of 1PI diagrams with which we work.  The application of the
Ward identity
to 1PI diagrams only, therefore, results in terms that would
cancel this special set of one-particle reducible diagrams,
in which the only line that spoils irreducibility is
contracted to a point.  These are the diagrams
shown in Fig.\ \ref{variationJ}, in which the ghost-gluon vertex of
Eq.\ (\ref{aBRS})
is inserted between one-particle irreducible  subdiagrams in all
possible ways.  The ghost line ending at this composite vertex is
continuously connected to the variation of a gluon propagator,
according to Eq.\ (\ref{varEquation}).
The full composite vertex of the Ward identity in Eq.\ (\ref{varEquation})
appears only at true external lines of the 1PI jet.  This vertex is given by
the momentum factor in Eq. (\ref{invPropag}) and is represented by 
the double line crossing a gluon line
in Fig. \ref{feynmanRulesF} below.
Diagrams that are reducible in one or more internal lines can be 
treated in a similar manner.
The ``left-over" terms in the Ward identities for each set of
diagrams of definite reducibility properties
(1PI, 2PI, etc.), must cancel in the full sum, reproducing the
identity for Green functions, Eq.\ (\ref{varEquation}).

\section{Tulip-Garden Formalism} \label{tulipGarden}

\begin{figure}
\includegraphics*{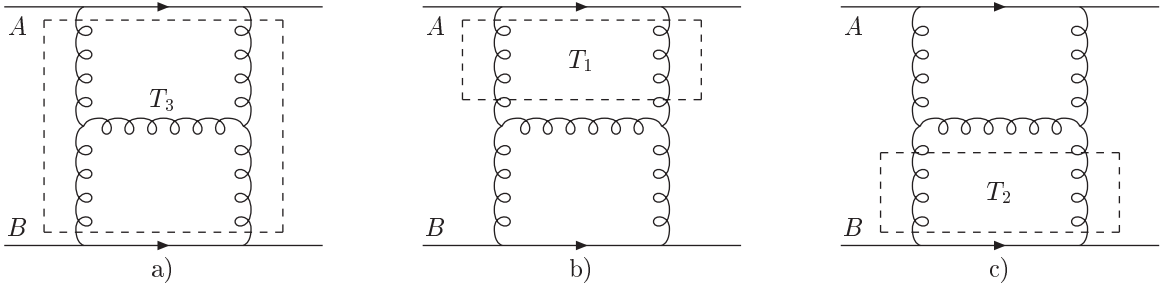}
\caption{ \label{tulip} Two-loop diagram illustrating the idea of
the tulip-garden formalism. $T_1, T_2, T_3$ are the possible tulips.}
\end{figure}

In this appendix we illustrate how a given Feynman diagram 
contributing to the process (\ref{qqqq}) in the
leading power can be systematically written in the form 
(\ref{fact1}). For concreteness let us consider a two
loop diagram where the quarks interact via the exchange of a one rung 
gluon ladder as in Fig. \ref{tulip}.
The important contributions of this diagram come from the regions 
when all of the exchanged gluons are soft,
Fig. \ref{tulip}a or when the gluons attached to the $A$ quark line 
are soft, while the rest of the gluons carries
momenta parallel to the $-$ direction (they belong to jet $B$), Fig. 
\ref{tulip}b, or when the two gluon lines
attached to the $B$ quark line are soft and the other gluons are 
collinear to the $+$
direction (they belong to jet $A$), Fig. \ref{tulip}c.
The possible central soft exchange parts are called tulips. In our 
case the possible tulips are denoted as
$T_1, T_2, T_3$ in Fig. \ref{tulip}. The garden is defined as a 
nested set of tulips
\{$T_1, \ldots, T_n$\} such that $T_{i} \subset T_{i+1}$ for $i=1, 
\ldots, n-1$.
In Fig. \ref{tulip}, $\{T_1\}$, $\{T_2\}$, $\{T_3\}$, $\{T_1, T_3\}$, 
$\{T_2, T_3\}$ are the possible gardens.

For a given tulip we make the soft approximation, consisting of 
attaching a soft gluon to
jet $A$ via the $-$ component of its polarization only and to jet $B$ 
via the the $+$ component of its polarization.
The result of this soft approximation for a given Feynman diagram $F$ 
corresponding to a tulip $T$ is denoted $S(T)F$.
It has obviously the form of Eq. (\ref{fact1}).
Following the prescription given in Refs. \cite{coso} and 
\cite{sen83} we write the contribution to a given diagram $F$ in the 
form
\be \label{tg}
F = \sum_{G} (-1)^{n+1} S(T_1) \ldots \, S(T_n) \, F + F_R,
\ee
where the sum over inequivalent gardens, as defined bellow, $G$ in 
Eq. (\ref{tg}) is understood.
The meaning of this expression is the following. For a given garden 
consisting of a set of
tulips $\{T_1, \ldots, T_n\}$, we start with the largest tulip $T_n$ 
and make the soft approximation
for the gluon lines coming out of it. Then for $T_{n-1}$ we proceed 
the same way as for $T_n$.
If some of the lines coming out of $T_{n-1}$ are identical to the 
ones coming out of $T_n$ we leave them untouched.
For instance, if we consider a garden $\{T_2, T_3\}$ from Fig. \ref{tulip},
we first perform the soft approximation on tulip $T_3$ and then 
proceed to tulip $T_2$.
However the lines coming out of $T_2$ and $T_3$ which attach to the 
$B$ quark line are identical
so when performing $S(T_2) S(T_3) F$ we leave these gluon lines out 
of the game and make soft
approximations only on the gluon lines attaching to the ladder's 
rung. Two gardens are equivalent
if the soft approximation is identical for both of them. $F_R$ is 
defined by Eq. (\ref{tg}).
The contribution to $F_R$ comes from the integration region where 
$|\vec{k}| \gtrsim \sqrt{s}$
for all gluons coming out of the central soft part. As a result, the 
contribution to $F_R$ is
suppressed by positive powers of $\sqrt{-t}/\sqrt{s}$.
Therefore we can ignore the contribution from $F_R$ within the 
accuracy at which we are working.

We can now rewrite Eq. (\ref{tg}),  as
\be
F = \sum_{T} \left( \sum_{G, T_n = T} (-1)^{n+1} \, S(T_1) \ldots \, 
S(T_{n-1}) \right)
     S(T) \, F + F_R.
\ee
This expression is indeed in the form of Eq. (\ref{fact1}) since the 
term $S(T) F$
is of that form and the subtractions $\sum (-1)^{n+1} S(T_1) \ldots 
\, S(T_{n-1})$ modify only the
soft function $S$ in Eq. (\ref{fact1}), but  do not alter the form of 
the equation.
We can therefore conclude that the contribution to a given Feynman 
diagram in leading power can be
expressed in the first factorized form given by Eq. (\ref{fact1}).

\section{Feynman Rules} \label{feynmanRules}

\begin{figure} \center
\includegraphics*{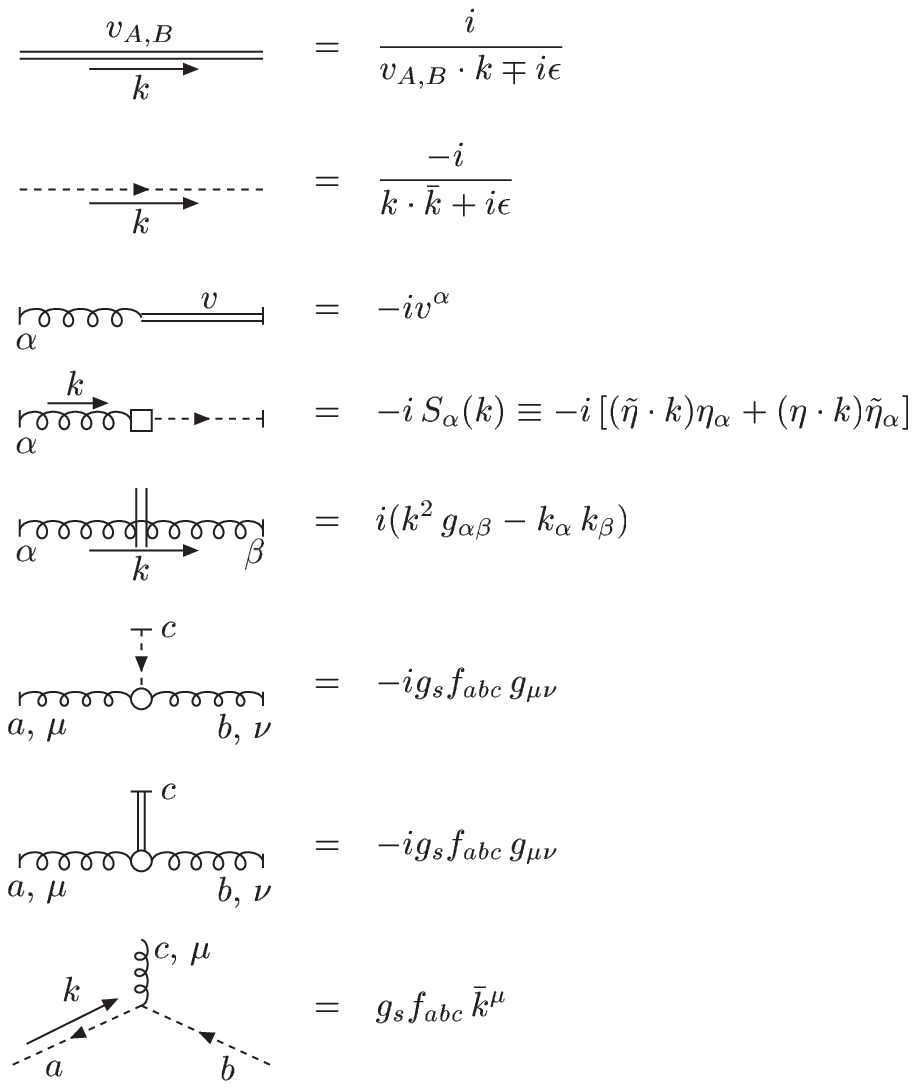}
\caption{ \label{feynmanRulesF} Feynman rules for the eikonal lines, 
ghost lines and special vertices.}
\end{figure}

In Fig. \ref{feynmanRulesF}, we list the Feynman rules for the lines 
and the vertices encountered in the text.
The double lines are eikonal lines, while the dashed lines represent 
ghosts. The four vectors $\eta, \, \tilde{\eta}$
are defined in Eqs. (\ref{eta}) and (\ref{etaTilde}), respectively.
The conventions for the gluon-ghost and gluon-eikonal vertices (third and 
second from the bottom of Fig. \ref{feynmanRulesF}) are the following.
We start with a color index of a gluon external to the diagram
defining the evolution kernel, see for instance Fig. \ref{trajll}a, 
then proceed to the gluon internal to the diagram and finally to the
ghost/eikonal line  in order to assign the color indices of 
$f_{abc}$. For the three point antighost - gluon - ghost
vertex at the bottom of Fig. \ref{feynmanRulesF}, 
we start with an antighost (arrow flowing out of the vertex) 
then proceed to the ghost and finally we reach the gluon line.

\section{Origin of Glauber Region} \label{glauberRegion}

\begin{figure} \center
\includegraphics*{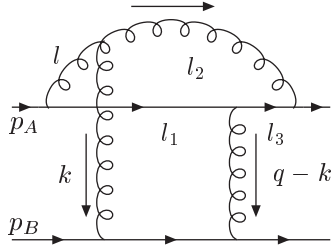}
\caption{\label{glauberExample} Two loop diagram demonstrating the 
origin of the Glauber (Coulomb) region.}
\end{figure}

In this appendix we exhibit the origin of the Glauber (Coulomb) region
using the two-loop diagram shown in Fig.
\ref{glauberExample}. Consider a situation when the upper gluon loop 
is a part of $J_A$. Momentum $k$ of the exchanged
gluon flows through jet lines  with momenta $l_2 = l - k$ and $l_3 = 
p_A - l - q + k$.
The components of $k$ can be pinched by
double poles coming from the denominators of the gluon propagators 
$k^2 + i\epsilon$ and $(q-k)^2 + i\epsilon$. In
addition to these pinches, the component $k^-$  can be pinched by the 
singularities of the jet lines $l_2$ and $l_3$,
at values
\bea \label{glauberPoles}
k^- & = & l^- - \frac{l_{2 \, \perp}^2 - i\epsilon}{2 l_2^+}, \nonumber \\
k^- & = & l^- + \frac{l_{3\perp}^2 - i\epsilon}{2 l_3^+}.
\eea
The two poles given by Eq. (\ref{glauberPoles}) are located in 
opposite half planes since in the region considered
$l_2^+, l_3^+ > 0$. This indicates that
we must consider the possibility that the
different components of the soft momentum $k$ can scale differently.
For instance, we can have  $k^+ \sim k_{\perp} \sim \sigma \sqrt{s}$ 
and $k^- \sim \lambda \sqrt{s}$ where $\lambda \ll \sigma \ll 1$.
Indeed, the power counting performed in Sec. \ref{elementary 
vertices} shows that the singularities originating
from these
regions can produce a logarithmic enhancement.  We also note that it 
is only minus components
that are pinched in this way by the lines in $J_A$, and plus 
components by the lines in $J_B$.

\end{document}